\documentclass{article}
\usepackage{emulateapj}
\usepackage{apjfonts}

\begin{document}

\newcommand{\FIGSTA}{1}
\newcommand{\FIGSTB}{2}
\newcommand{\FIGTA}{3}
\newcommand{\FIGTB}{4}
\newcommand{\FIGFLA}{5}
\newcommand{\FIGC}{6} 
\newcommand{\FIGFLB}{7}
\newcommand{\FIGINT}{8}
\newcommand{\FIGR}{9}
\newcommand{\FIGCOMA}{10}
\newcommand{\FIGCOMB}{11}
\newcommand{\FIGCTHB}{12} 
\newcommand{\FIGMETA}{13}  
\newcommand{\FIGCTHC}{14} 
\newcommand{\FIGMETC}{15}  
\newcommand{\FIGMETB}{16}  
\newcommand{\FIGIONA}{17} 
\newcommand{\FIGIONB}{18} 
\newcommand{\FIGCONV}{19}

\newcommand{\lta}{\stackrel{<}{\sim}}
\newcommand{\gta}{\stackrel{>}{\sim}}

\setcounter{figure}{1}

\submitted{To appear in The Astrophysical Journal}

\title{Non-LTE Models and Theoretical Spectra of Accretion Disks
in Active Galactic Nuclei.  
IV. Effects of Compton Scattering and Metal Opacities}
 
\author{Ivan Hubeny}
\affil{AURA/NOAO, NASA Goddard Space Flight Center, Code 681, Greenbelt, 
       MD 20771.  hubeny@tlusty.gsfc.nasa.gov}

\author{Omer Blaes}
\affil{Department of Physics, University of California, Santa Barbara,
CA 93106.   blaes@gemini.physics.ucsb.edu}

\author{Julian H. Krolik}
\affil{Department of Physics and Astronomy, Johns Hopkins University,
           Baltimore, MD 21218.   jhk@ tarkus.pha.jhu.edu}

\and
\author{Eric Agol\altaffilmark{1}}
\affil{Department of Physics and Astronomy, Johns Hopkins University,
           Baltimore, MD 21218.   agol@woody.tapir.caltech.edu}
\altaffiltext{1}{present address: California Institute of Technology,
    Pasadena, CA 91125}

\begin{abstract}
We extend our models of the vertical structure and emergent
radiation field of accretion disks around supermassive black holes
described in previous papers of this series.  Our models now include
both a self-consistent treatment of Compton scattering and the
effects of continuum opacities of the most important metal species (C, N,
O, Ne, Mg, Si, S, Ar, Ca, Fe, Ni).  With these new effects
incorporated, we compute the predicted spectrum from black holes accreting
at nearly the Eddington luminosity ($L/{\rm L}_{\rm Edd}\approx0.3$)
and central masses of $10^6$, $10^7$, and $10^8 M_{\odot}$.  We
also consider two values of the Shakura-Sunyaev $\alpha$ parameter,
0.1 and 0.01, but in contrast to our previous papers, we consider
a kinematic viscosity which is independent of depth. 

Although it has little effect when
$M > 10^8 M_{\odot}$, Comptonization grows in importance as the central
mass decreases and the central temperature rises.  
It generally produces an increase
in temperature with height in the uppermost layers of hot atmospheres.
Compared to models with coherent electron scattering,
Comptonized models have enhanced extreme ultraviolet/soft X-ray emission,
but they also have a more sharply declining spectrum at very high frequencies.
Comptonization also smears the hydrogen and the He~II Lyman edges.  
The effects of metals on the overall spectral energy distribution 
are smaller than the effects of Comptonization for
these parameters.  Compared to pure hydrogen-helium models, models
with metal continuum opacities have reduced flux in the high frequency tail,
except at the highest frequencies, where the flux is very low.
Metal photoionization edges are not present in the
overall disk-integrated model spectra.

    The viscosity parameter $\alpha$ has a more dramatic effect on
the emergent spectrum than do metal continuum opacities.  As
$\alpha$ increases (and therefore the disk column density decreases),
the flux at both the high and low frequency extremes of the spectrum
increases, while the flux near the peak decreases.
Multitemperature blackbodies are a very poor approximation to accretion disk
spectra in the soft X-ray region, and such crude modeling may greatly
overestimate the accretion luminosity required to explain observed soft
X-ray excesses in active galactic nuclei.
In addition to our new grid of models, we also present a simple
analytic prescription for the vertical temperature structure of the disk
in the presence of Comptonization, and show under what conditions
a hot outer layer (a corona) is formed.
\end{abstract}


\section{Introduction}

In previous papers of this series (Hubeny \& Hubeny 1997, 1998; Hubeny et al.
2000; hereafter referred to as Papers I, II, and III, respectively),
we computed a series of models of accretion disks around supermassive 
black holes appropriate for quasars and active galactic nuclei (AGN). 
The models included a set of 1-D vertical structure calculations for the 
individual annuli of the disk.  We solved self-consistently the equations
of vertical hydrostatic equilibrium, energy balance, radiative transfer, 
and, since we did not assume local thermodynamic equilibrium (LTE), 
the set of statistical equilibrium equations for selected energy 
levels of various species of atoms and ions.
In these previous models, we considered a simple H-He chemical composition, 
and we assumed that electron scattering was coherent (i.e., Thomson 
scattering).

In the present paper, we refine the previous modeling procedure by relaxing
these two limiting approximations.  We now incorporate metals, and we include
the effects of incoherent electron scattering, i.e. Compton scattering.
Both of these effects are expected to modify the emergent spectrum of 
individual hot annuli which emit predominantly in the extreme ultraviolet
and soft X-rays.  Excess emission above a power law extending down from the
hard X-rays is widely observed in many classes of AGN, from Seyferts to quasars
(e.g. Turner \& Pounds 1989).  The origin of this emission
remains unclear, but it has been proposed to be thermal, Comptonized emission
from hot portions of an accretion disk (e.g. Czerny \& Elvis 1987; Ross,
Fabian \& Mineshige 1992).  In addition, fluorescent line emission and
Compton reflection humps are observed in AGN (Nandra \& Pounds 1994)
and appear to be at least
partly the result of external illumination of the disk by hard X-rays from a
hot, coronal region somewhere in the vicinity.
The effects of Compton scattering and metal opacities within the disk itself
are crucial for understanding this phenomenology.  Moreover, these effects
are still more pronounced in the even higher temperature accretion disks
thought to exist around stellar mass black hole candidates in X-ray binaries.

This paper is organized as follows.  In section~2 we describe in detail our
basic physical assumptions and numerical methods in treating the effects
of Comptonization and metal opacities.  We then describe in section~3 an
approximate analytic calculation of the vertical temperature structure of
an atmosphere which is subject to Comptonization.  In section~4 we present
our new numerical models of hot accretion disks, including a detailed
comparison with our previous work and that of other authors.  
In section~5 we discuss these results and then we summarize 
our conclusions in section~6.

\section{Method}

The basic equations for the vertical structure were discussed in detail 
in Paper~II, to which the reader is referred for more information. 
The main difference here is an inclusion of Compton scattering. 
Since our treatment differs somewhat from more commonly used approaches
(e.g., Ross, Fabian, \& Mineshige 1992; Shimura \& Takahara 1993
in the context of AGN disks; or Madej 1989, 1991, 1998 in the context
of model atmospheres of neutron stars and white dwarfs), 
we will describe it here in more detail.


\subsection{Treatment of Compton Scattering}

A general treatment of Compton scattering of (unpolarized)
radiation is described in Pomraning (1973).  The effects of Compton
scattering are described through an appropriate frequency- and
angle-dependent redistribution function that incorporates
Klein-Nishina corrections and the relativistic Maxwellian
velocity distribution of the electrons.  In the context of stellar
atmosphere modeling, this approach was used e.g. by Madej (1989;
1991, 1998; Madej \& Rozanska 2000) for computing model
atmospheres of neutron stars, white dwarfs, and X-ray irradiated
B stars.

We are interested here in accurately treating the extreme ultraviolet/soft X-ray
emission of annuli with temperatures much less than the electron rest mass
energy.  In this non-relativistic regime, Compton scattering can be
incorporated in the radiative transfer equation through a Kompaneets-like
term, i.e. by replacing the exact angle-frequency redistribution integral
by a Fokker-Planck form.  We present details of how this can be implemented
numerically in Appendix A.  For all the calculations presented in this paper,
we use a Compton source function which is averaged over all angles,
analogous to the standard isotropic Kompaneets equation (e.g. Rybicki \&
Lightman 1979).  We stress, however, that we do not assume that the radiation
field is isotropic.  Once an atmosphere structure has converged, our treatment
could be improved by performing a formal solution of the radiative transfer
equation with the full angle-dependent Compton source function (see Appendix
A).

We adopt the notation
commonly used in stellar atmosphere modeling, i.e. we express a photon's
energy through its frequency, $\nu$, and the photon occupation number
through the specific intensity of radiation, $I_\nu$. Also, we often formally
express a dependence on frequency through a subscript $\nu$.
In the plane-parallel atmospheres we are considering, the radiative transfer
equation may be written as
\begin{equation}
\label{rte}
\mu {\partial I_\nu (\mu)\over \partial \tau_\nu} =
I_\nu (\mu) - \epsilon_\nu S_\nu^{\rm th} - \lambda_\nu
\bar S_\nu^{\rm Compt} \, .
\end{equation}
Here $\mu$ is the direction cosine between the photon propagation direction
and the vertical.  The monochromatic optical depth, $\tau_\nu$, is defined by
\begin{equation}
d\tau_\nu = - (\kappa_\nu^{\rm th} + \sigma_\nu)\, dz 
\equiv - \chi_\nu \, dz\,  ,
\end{equation}
where $\kappa_\nu^{\rm th}$ is the total thermal absorption coefficient at
frequency 
$\nu$ (including contributions from all bound-free, free-free, and 
possibly bound-bound transitions which operate at this frequency), and
the scattering coefficient is given by
\begin{equation}
\label{eqsig}
\sigma_\nu = n_e \sigma_{\rm T} \, (1 - 2 x).
\end{equation}
Here $\sigma_{\rm T}$ is the Thomson cross-section, and
the dimensionless frequency $x$ is given by
\begin{equation}
x = {h\nu \over m_e c^2}\, .
\end{equation}
The $-2x$ term in equation (\ref{eqsig}) accounts for the Klein-Nishina
reduction in scattering cross-section, to first order in $x$.

We define the ``photon destruction probability'', $\epsilon_\nu$, as
\begin{equation}
\epsilon_\nu =  {\kappa_\nu^{\rm th} \over \kappa_\nu^{\rm th} + 
\sigma_\nu }\, .
\end{equation}
and the ``scattering probability'', $\lambda_\nu$, as
\begin{equation}
\lambda_\nu =  {n_e \sigma_{\rm T} \over \kappa_\nu^{\rm th} + \sigma_\nu }
= {1-\epsilon_\nu\over 1-2x} \, .
\end{equation}
Note that in contrast to the case of coherent scattering,
$\lambda_\nu \neq 1-\epsilon_\nu$.
The thermal source function is given by
\begin{equation}
S_\nu^{\rm th} = {\eta_\nu^{\rm th} \over \kappa_\nu^{\rm th}}\ ,
\end{equation}
where $\eta_\nu^{\rm th}$ is the thermal emission coefficient, 
which contains contributions from all the spontaneous inverse processes 
to those giving rise to thermal absorption at frequency $\nu$. 
As is customary, we treat the stimulated emission for these processes
as negative absorption.

The angle-averaged ``Compton source function'' is given by
\begin{equation}
\label{compt_sf}
\bar S_\nu^{\rm Compt} =
(1-x) J_\nu + (x-3\Theta)  J_\nu^\prime + \Theta J_\nu^{\prime\prime} +
{c^2 \over 2 h \nu^3} J_\nu 2x (J_\nu^\prime - J_\nu)\, ,
\end{equation}
where
\begin{equation}
\Theta = {kT \over m_e c^2}\, ,
\end{equation}
$J_\nu$ is the angle-averaged intensity, and $J_\nu^\prime$ and
$J_\nu^{\prime\prime}$ are defined as
\begin{equation}
J_\nu^\prime = {\partial J_\nu\over \partial \ln \nu}\, 
\,\,\,\,{\rm and}\,\,\,\,
J_\nu^{\prime\prime} = {\partial^2 J_\nu\over \partial (\ln \nu)^2}\, .
\end{equation}
A detailed derivation of equation (8) is presented in Appendix A, but we note
that it is simply equivalent to taking the emission terms of the isotropic
Kompaneets equation (e.g. Rybicki \& Lightman 1979), and replacing the photon
occupation number with $J_\nu/\nu^3$.
The last term on the right-hand side of equation (\ref{compt_sf}) represents
stimulated Compton scattering.  Note that the Thomson limit of the scattering
source function can be formally obtained by setting $x=\Theta=0$ in equation
(\ref{compt_sf}), in which case we obtain
$ S_\nu^{\rm Compt} = J_\nu$, i.e., the usual Thomson scattering source
function for isotropic scattering.

We write the energy balance equation in a manner very similar to equation (36) 
of Paper~II, viz.
\begin{equation}
\label{energy}
{\partial F_z \over \partial z}  =
4\pi \int_0^\infty \chi_\nu (S_\nu^{\rm tot} - J_\nu )d\nu\, .
\end{equation}
Here $F_z$ is the vertical component of the energy flux. 
Note that in the case of coherent (Thomson) scattering, the absorption and
emission contributions exactly cancel in the right hand side of equation (11),
so the net radiative cooling term contains only thermal absorption and
emission terms.  Compton scattering, however, affects the thermal
balance, so we have to retain the {\it total} absorption coefficient $\chi_\nu$
and the {\em total} source function, $S_\nu^{\rm tot}\equiv\epsilon_\nu
S_\nu^{\rm th}+\lambda_\nu\bar{S}_\nu^{\rm Compt}$.

The coupled set of structural equations (\ref{rte}), (\ref{energy}), 
together with the hydrostatic equilibrium and statistical equilibrium 
equations (given explicitly in Paper~II), discretized in frequency and depth, 
is solved by the hybrid Complete Linearization/Accelerated Lambda Iteration 
(CL/ALI) method (Hubeny \& Lanz 1995).  
The only new ingredient here, the Compton source function, is given by
\begin{equation}
\label{compt_discr}
S_i^{\rm Compt} = A_i J_{i-1} + B_i J_i + C_i J_{i+1} +
                  J_i (U_i J_{i-1} + V_i J_i + W_i J_{i+1} )\ ,
\end{equation}
where $i$ is the frequency index, $J_i \equiv J(\nu_i)$ , 
and the coefficients $A$ - $C$ and
$U$ - $W$ are functions of the parameters $x$ and $\Theta$. They also 
depend on the form of the
difference formula for differential operators in frequency.
We use here a simple second order difference formula.
Detailed expressions for the coefficients are given in Appendix A.
Equation (\ref{compt_discr})
is valid for the interior points of the frequency grid: $i = 2, \ldots, NF-1$
($NF$ being the total 
number of frequency points). For boundary conditions in frequency 
space, we use the following procedure: \
(i) For the lowest frequency ($i=1$), we assume that the mean intensity follows
the Planck function in the Rayleigh-Jeans limit,
$J_\nu \propto (2kT/c^2) \nu^2$
so that
\begin{equation}
\label{jlbound}
J_\nu^\prime = 2 J_\nu\, , \quad\quad J_\nu^{\prime\prime} = 4 J_\nu\, .
\end{equation}
(ii) Analogously, for the highest frequency ($i=NF$), we assume the Wien limit,
\begin{equation}
\label{jubound}
J_\nu \propto {2 h\nu^3\over c^2} \exp(-h\nu/kT)\, ,
\end{equation}
and thus
\begin{equation}
J_\nu^\prime = (3-z) J_\nu\, , \quad\quad 
J_\nu^{\prime\prime} = [(3-z)^2 - z] J_\nu\, ,
\end{equation}
where
\begin{equation}
z= h\nu/kT = x/\Theta \, .
\end{equation}
The off-diagonal terms $A_i$, $C_i$, $U_i$, and $W_i$ 
therefore vanish for $i=1$ and $i=NF$.

Since the boundary condition (\ref{jubound}) is somewhat artificial,
we have also tested another type of boundary condition, assuming
that the mean intensity of radiation at the highest frequency is
given by a power-law,
\begin{equation}
\label{jupl}
J_\nu \propto \nu^{-\beta}\, ,
\end{equation}
in which case
\begin{equation}
J_\nu^\prime = -\beta J_\nu\, , \quad\quad 
J_\nu^{\prime\prime} = \beta^2 J_\nu\, .
\end{equation}
We have found that the choice of boundary conditions only marginally
influences the radiation intensity.  This is because we adopt the highest
frequency to be so high that the emergent flux is lower than the flux at
the peak by several orders of magnitude.  

The hybrid CL/ALI scheme allows for an arbitrary partitioning of ``linearized'' 
and ``ALI'' frequency points, i.e., those where the mean intensity of 
radiation is linearized, and those treated through the ALI scheme 
(and therefore not linearized), respectively. 
For all the calculations in this paper, we chose frequencies 
in the continuum with $\nu > 10^{15}$ s$^{-1}$ to be linearized, 
while the rest were treated within the ALI scheme.  Hence all the frequencies
where Compton scattering could be important were linearized.
In the future, we intend to implement an improved ALI treatment of 
Comptonization that will allow us to treat all frequencies as ALI frequencies, 
which would lead to a significant reduction of computer time. Work on this 
problem is in progress and will be described in a future paper.

In the linearized transfer equation, there is no problem treating the
frequency coupling induced by Comptonization (i.e., the presence of the 
$A_i$ and $C_i$ terms in equation \ref{compt_discr})
or the non-linearity due to stimulated Compton scattering.  
However, before entering a linearization step, one must perform  
a ``formal solution'' of the transfer equation, i.e., a solution
with the known, current thermal source function. This is very easy 
in the case of Thomson scattering because one can simply solve the transfer 
equation for one frequency at a time. The angular dependence of the radiation 
field is also easy to deal with using the Feautrier method (e.g., Mihalas 1978). 
The formal solution determines the variable Eddington factor,
$f_\nu \equiv K_\nu/J_\nu$, where $K_\nu$ is the second moment of the
specific intensity: $K_\nu = (1/2) \int_{-1}^{1} I_\nu(\mu) d\mu$.

In the case of Compton scattering, even the formal solution presents numerical
complications because we cannot use the frequency-by-frequency strategy.  
In the present version of our computer program TLUSDISK (see also Paper~II),
we use the following procedure for the formal solution of the transfer equation
with Comptonization.
We first solve the transfer equation in the second-order form
(e.g., Mihalas 1978), viz.
\begin{equation}
\label{rte_ii}
{\partial^2 (f_i J_i) \over \partial \tau_i^2} = 
J_i -S_i^{\rm tot}\, ,
\end{equation}
where we assume a current value of the Eddington factor $f_i$.
We then discretize in depth, and organize the mean intensities $J_{i,d}$ ($i$ 
is the frequency index and 
$d=1,\ldots, D$ is the depth index) in a set of column vectors
\begin{equation}
\label{jvec_def1}
{\bf J}_i \equiv \left( J_{i,1}, J_{i,2}, \ldots, J_{i,D} \right)^T \, ,
\end{equation}
so that the mean intensity vector contains intensities at all depth points
for a given frequency point $i$. The resulting discretized transfer equation
reads
\begin{equation}
\label{rte_mat1}
- {\bf A}_i {\bf J}_{i-1} + {\bf B}_i {\bf J}_i - {\bf C}_i {\bf J}_{i+1} =
{\bf L}_i\, .
\end{equation}
Here the matrices ${\bf B}$ are tridiagonal (because of the difference 
representations of the second derivative with respect to depth), 
while the matrices ${\bf A}$ and ${\bf C}$ are diagonal (because the terms
containing the frequency derivatives are local in physical space).
Detailed expressions for the elements of matrices ${\bf A}$, ${\bf B}$ 
and ${\bf C}$ are given in Appendix A.
We avoid the nonlinearity due to the stimulated scattering term by
writing the Compton source function as
\begin{equation}
\label{compt_discra}
S_i^{\rm Compt} = A_i J_{i-1} + B_i J_i + C_i J_{i+1} +
                  J_i (U_i J_{i-1}^{\rm old} + V_i J_i^{\rm old} + 
		  W_i J_{i+1}^{\rm old} )\, ,
\end{equation}
i.e., by assuming that the stimulated contribution is evaluated 
using the current (``old'') values of the mean intensity. 
If needed, the formal solution may be iterated to obtain an improved 
stimulated contribution, or one may devise a linearization
procedure for the formal solution itself (which is distinct from 
our global linearization where we solve simultaneously all the 
structural equations).  We have tested both approaches, and found that such 
refinements are never needed in actual model calculations.

Solution of equation (\ref{rte_mat1}) is done by
standard Gauss-Jordan elimination, consisting of a forward elimination 
followed by a back-substitution, just as in the original Feautrier
method for the Thomson scattering case. 

Finally, the new Eddington factors are obtained by performing a set of
elementary formal solutions for one frequency and angle at a time, with the
total source function evaluated using the previously determined values of the
mean intensity at all frequencies. 


\subsection{Treatment of Metals}

We consider the following chemical species: H, He, C, N, O, Ne, Mg, Si, S, Ar,
Ca, Fe, and Ni.  We allow for all possible ionization stages, from neutrals
to fully stripped atoms (e.g., Fe~I to Fe~XXVII).  For the
computations presented in this paper, we treat all ions except H~I and
He~II as one-level atoms. This is done in order to explore the basic
effects of metal opacities, together with Comptonization and departures
from LTE, in an efficient way.  We plan to use more extended model atoms
in future calculations; very likely by adopting model atoms similar to those
in the X-ray photoionization code XSTAR (Kallman 2000).  For hydrogen, we
use a 9-level model atom (where the ninth level is a merged ``superlevel''
of all higher states) as in Papers~I-III.  He~II is represented as a
14-level atom. Since we consider only hot disks here, He I is represented
by a 1-level atom, in contrast to Papers~I-III where it was represented by
a 14-level atom.  We tested that this approximation does not lead to
any errors since He I has negligible abundance in the current models.

For hydrogenic ions, we use the standard analytic expression for the
photoionization cross-section.  
For all nonhydrogenic metal ion photoionization
cross-sections, we use data from XSTAR, version 1, kindly supplied by
Tim Kallman (subroutine {\tt bkhsgo}, based on data from Barfield, Koontz,
\& Huebner 1972).  XSTAR was recently upgraded to version 2, and now
uses more recent sources of photoionization cross-sections (the Opacity Project
and others); we also plan to adopt these cross-sections in future
calculations.

We take into account all inner-shell photoionization processes, but make
the simplifying assumption that if an Auger electron is energetically
possible, then it is in fact produced and the photoionization results
in a jump by two stages of ionization to a ground state configuration.
We therefore neglect both fluorescence and multiple Auger electron ejection
arising from inner shell photoionization.  Note that this is not a good
approximation for low ionization stages of metals, where multiple Auger
electron ejection is likely after an inner shell photoionization (e.g.
Kaastra \& Mewe 1993).  This might be important in regions where non-LTE
effects are important in determining the abundance of these ion species.
In future calculations we will lift these simplifying assumptions and
consider more elaborate inner shell photoionization branching possibilities.

We include the effects of collisional ionization of the outermost valence
shell electron of each metal ion species by again using data from XSTAR,
version~1 (subroutine {\tt cion}).
Once again, we plan on
improving this treatment with more modern data in the future.

Dielectronic recombination is handled using the subroutine {\tt direc} from
XSTAR, version~1.  This is based on data from Aldrovandi \& Pequignot (1973),
Nussbaumer \& Storey (1983), and Arnaud \& Raymond (1992).  
Numerically, we treat dielectronic recombination by introducing an
artificial modification of the photo-ionization cross-section, which has
the advantage that the detailed balance relations are automatically
satisfied. We describe this procedure in more detail in Appendix B.



\section{Understanding Temperature Structure in the Presence of
Comptonization}

Before we turn to actual numerical results, it is worthwhile developing
a simple analytic model that will allow us to understand
the vertical temperature profile in the presence of Compton
scattering. A similar study for the case of ``classical'' disks without
Comptonization was done by Hubeny (1990).  The approach developed here
is its generalization.
We note that Madej (1989) considered a similar LTE-gray
model in the context of neutron star atmospheres (i.e., with
no viscous heating), although he only provided a numerical solution.


\subsection{Energy Balance}

The energy balance equation may be written in the form (see Paper~II)
\begin{equation}
\label{ener3}
{d F_{\rm rad} \over dm } \, = \, 
- \sigma T_{\rm eff}^4 \, {w(m) \over m_0 \bar w}\, .
\end{equation}
Here $w(m)$ is the dissipation rate at column density $m$ and $\bar w$
is the vertical mean of $w$.  In evaluating $w$ we employ the relativistic
corrections of Riffert \& Herold (1995) as in Paper~II; replacing them
with the forms advocated by Abramowicz et al. (1997) produces no significant
change in the emergent spectrum.

The important point is that we can express the gradient of the
radiation flux through the transfer equation
\begin{equation}
\label{ener3a}
{d F_{\rm rad} \over dm } \, = \, 
4\pi \int_0^\infty \chi_\nu (S_\nu^{\rm tot} - J_\nu )d\nu \, .
\end{equation}
Equating now the right-hand sides of equations (\ref{ener3}) and
(\ref{ener3a}) gives another form of the energy balance equation, which
expresses the fact that the total energy generated locally is
balanced by the net local radiative loss.

The integrand of the radiative loss term is given by
\begin{eqnarray}
\label{radloss}
\chi_\nu (S_\nu^{\rm tot} - J_\nu ) =
\kappa_\nu \left(S_\nu^{\rm th} - J_\nu \right) +
\quad\quad\quad\quad \quad\quad\quad  \nonumber \\
 n_{\rm e}\sigma_{\rm T}
\left[ x J_\nu + (x-3\Theta)  J_\nu^\prime + \Theta J_\nu^{\prime\prime} +
{c^2 \over 2 h \nu^3} J_\nu 2x (J_\nu^\prime - J_\nu) \right]\, .
\end{eqnarray}
In the following, we neglect the stimulated scattering term, and assume
that the thermal source function is given by the Planck function.
We stress that this is done in this section only for developing a
simple analytic model; in actual calculations these simplifications
are not made.  The integral in Eq. (\ref{ener3a}) can
be evaluated using integration by parts for the derivative terms.
After some algebra we obtain
\begin{equation}
\label{ener4}
\kappa_B B - \kappa_J J + \kappa_\sigma (4\Theta- \bar x) J = 
 {\sigma\over 4\pi} \, T_{\rm eff}^4 \, {w(m) \over m_0 \bar w}\, ,
\end{equation}
where $J=\int_0^\infty J_\nu d\nu$, and analogously for $B$.  The
mean opacities are defined by
\begin{equation}
\kappa_J = \int_0^\infty (\kappa_\nu^{\rm th}/\rho) J_\nu d\nu/ J\, ,
\end{equation}
\begin{equation}
\kappa_B = \int_0^\infty (\kappa_\nu^{\rm th}/\rho) B_\nu d\nu/ B\, ,
\end{equation}
(which is called the Planck-mean opacity), and
\begin{equation}
\kappa_\sigma = n_{\rm e}\sigma_{\rm T}/\rho \, .
\end{equation}
Finally, $\bar x$ is defined by
\begin{equation}
\bar x = {h \over mc^2} \, \bar\nu =  {h \over mc^2} \,\, 
          {\int_0^\infty \nu J_\nu d\nu \over J}\, ,
\end{equation}
which has the meaning of a mean frequency weighted by the local
mean intensity.	
Since the mean intensity is typically a sharply-peaked function of frequency,
$\bar\nu$ is close to the frequency where $J_\nu$ attains its maximum value.
We note that the traditional {\em Compton temperature}, $T_C$, is defined
(e.g., Krolik 1999) as the temperature for which the term
$4\Theta- \bar x$ vanishes, i.e.
\begin{equation}
T_C = {h \bar \nu\over 4 k} \approx 1.2\times 10^{-11}\, \bar\nu \, .
\end{equation}

Equation (\ref{ener4}) has a simple physical interpretation.
The right-hand side expresses the net energy generated per unit mass
at a given depth in the disk. The left-hand side expresses the net energy
loss, i.e. the energy that has been transferred from material particles
to photons. The term $\kappa_B B - \kappa_J J$ expresses the transfer
of energy from particles to photons by {\em thermal processes}
(i.e. bound-bound, bound-free, and free-free atomic processes),
while the term $\kappa_\sigma (4\Theta- \bar x) J$ expresses the
net loss of energy of particles, in this case electrons, by
Compton scattering.  The latter term is in fact a generalization of the
usual formula (e.g., Rybicki \& Lightman 1979) for the net photon energy
gain (i.e., net electron energy loss) in a single scattering, 
which in our notation reads
\begin{equation}
{\Delta(h \nu) \over h\nu} = 4 \Theta - x\, .
\end{equation}
We also note that the ratio $\kappa_B/\kappa_\sigma$ can be expressed as
\begin{equation}
\label{epsbar}
\kappa_B/\kappa_\sigma = \int_0^\infty (\kappa_\nu^{\rm th}/(n_{\rm e}
\sigma_{\rm T})
B_\nu d\nu/ B \approx \int_0^\infty \epsilon_\nu B_\nu d\nu/B \equiv
\bar\epsilon \, ,
\end{equation}
i.e. the ratio of the Planck-mean to the scattering opacity has the
meaning of a {\it mean photon destruction parameter}.
Notice that the last approximate equality only holds in the case of
dominant electron scattering, $\epsilon_\nu \ll 1$.

To proceed further, we need an expression for the local integrated
mean intensity, $J$. This can be obtained from the second moment of the
radiative transfer equation 
\begin{equation}
\label{rtem2}
{d K_\nu \over dm} =  {\chi_\nu \over \rho}\, H_\nu  \, ,
\end{equation}
where $H_\nu= (1/2) \int_{-1}^1 I_\nu(\mu) d\mu$ is the first moment of
the specific intensity.
We now invoke the Eddington approximation, i.e.,
$K_\nu= J_\nu/3$, and integrate equation (\ref{rtem2})
over all frequencies.  We obtain, again in the case of dominant electron
scattering,
\begin{equation}
\label{jinteg}
{d J \over dm} \approx 3 \kappa_\sigma H = {3\kappa_\sigma\over 4\pi} 
F_{\rm rad}\, .
\end{equation}
The local radiation flux follows from integrating equation (\ref{ener3}), viz.
\begin{equation}
\label{fradm}
F_{\rm rad}(m) = \sigma\, T_{\rm eff}^4 
\left[ 1- {\int_0^m w(m^\prime) dm^\prime \over  m_0 \bar w} \right ] \, .
\end{equation}
Substituting equation (\ref{fradm}) into (\ref{jinteg}) and solving it for
$J$ we obtain
\begin{equation}
\label{jm}
J(m)= {3\sigma\over 4\pi}T_{\rm eff}^4 \, \left( {1\over \sqrt{3}} +
\tau - \tau_w \right) \, ,
\end{equation}
where we used the Eddington approximation form of the boundary condition,
$J(0) = H(0) \sqrt{3}$, and expressed $H(0)$ through the effective temperature.
The optical depth is defined by
\begin{equation}
\tau = \int_0^m \kappa_\sigma dm^\prime \approx \kappa_\sigma m\,
\end{equation}
since we assume that hydrogen and helium are completely ionized so that
$\kappa_\sigma$ is nearly constant. 
This optical depth is thus the Thomson optical depth.
The other optical depth in equation (\ref{jm}) may be called the
``viscosity-weighted'' optical depth, and is defined by
\begin{equation}
\tau_w = \int_0^m dm^\prime \kappa_\sigma  
         \int_0^{m^\prime} w(m^{\prime\prime}) dm^{\prime\prime}\, 
	 /m_0 \bar w \, .
\end{equation}
In the case of a depth-independent kinematic viscosity,
\begin{equation}
\tau_w =  {\tau^2\over 2\tau_{\rm tot}} 
\end{equation}
where $\tau_{\rm tot} \equiv \kappa_\sigma m_0$ is the Thomson optical depth
at the midplane.

Finally, combining equations (\ref{ener4}) and (\ref{jm}), we arrive
at the basic equation of this section,
\begin{equation}
\label{ener_bas}
\kappa_B \left({T \over T_{\rm eff} } \right)^4 = {3\over 4} 
\left( {1\over\sqrt{3}} +\tau - \tau_w \right) \left[\kappa_J - \kappa_\sigma
(4\Theta-\bar x) \right] + {w(m) \over 4 \bar w m_0}  \, ,
\end{equation}
where we used the well-known expression $B = (\sigma/\pi) T^4$.

Using equation (\ref{epsbar}), and assuming, as usual, that 
$\kappa_J = \kappa_B$, we rewrite equation (\ref{ener_bas}) in
a more instructive form,
\begin{equation}
\label{tempg1}
\bar\epsilon \left({T \over T_{\rm eff} } \right)^4 = {3\over 4} 
\left( {1\over\sqrt{3}} +\tau - \tau_w \right) (\bar\epsilon 
-4\Theta+\bar x)  + {w(m) \over \bar w} {1\over 4\tau_{\rm tot}} \, .
\end{equation}
We see that the effects of Compton scattering are negligible if
$| 4\Theta-\bar x | \ll \bar\epsilon$, in which case we
recover equation 3.11 of Hubeny (1990).

Before we discuss the meaning and implications of 
equation (\ref{tempg1}) in more detail, we first turn to
estimating the mean frequency.


\subsection{The Mean Frequency}

A determination of the mean frequency is easy if the mean intensity
is proportional to the Planck function, $J_\nu = a B_\nu$, where
$a$ is a frequency-independent constant.  In this case
\begin{equation}
\label{nubar2}
\bar\nu = {\int_0^\infty \nu^4/[\exp(h\nu/kT)-1]\, d\nu \over
\int_0^\infty \nu^3/[\exp(h\nu/kT)-1]\, d\nu} = 
{kT\over h}\,  {\Gamma(5)\over\Gamma(4)}\, {\zeta(5)\over \zeta(4)}
\approx 3.83 \, {kT\over h}\, .
\end{equation}
Here, $\Gamma$ is the Euler gamma function and $\zeta$ is the
Riemann zeta function (e.g., Abramowitz \& Stegun 1970).
The Compton cooling term is thus given by
\begin{equation}
\label{comcool}
4\Theta - \bar x = \Theta \, (4 - 3.83) = 0.17\, \Theta 
= 2.867 \times 10^{-11}\, T\, .
\end{equation}
Deep in the atmosphere, $J_\nu \approx B_\nu(T)$, where $T$ is the
local temperature. Therefore, the Compton cooling term is {\em positive},
and is determined by the {\em local} temperature.

At the surface, the mean frequency can be estimated as follows.
Let us assume for simplicity 
that the total Thomson optical depth is large. Then the surface
mean intensity, assuming the Eddington approximation, is given by
(see, e.g., Mihalas 1978 or Rybicki \& Lightman 1979)
\begin{equation}
\label{jsurf1}
J_\nu(\tau=0) \approx \sqrt{\epsilon_\nu}\,  
B_\nu(T(\tau=1/\sqrt{3\epsilon_\nu}))\, .
\end{equation}
In other words, the mean intensity is still proportional to the
Planck function; however the constant of proportionality, 
$\sqrt{\epsilon_\nu}$, depends on frequency, and the temperature
at which the Planck function is evaluated is {\em not} the local
temperature, but the temperature at depth 
$\tau\equiv\tau_\nu^{\rm form}=1/\sqrt{3\epsilon_\nu}$,
which may actually be quite large.
We call the depth $\tau_\nu^{\rm form}$ the {\em effective depth of
formation}.  We note that there is an intimate relation between the
effective depth of formation and the traditional {\em effective
optical depth} (e.g., Rybicki \&~ Lightman 1979; their equation 1.98).
The effective optical depth is given by (in our notation)
\begin{equation}
\tau_\nu^\ast \approx \sqrt{\tau_\nu^{\rm th}\, \tau_\nu} 
\approx \tau_\nu \, \sqrt{\epsilon_\nu}\, .
\end{equation}
Our effective depth of formation is thus equal, up to a factor of the
order of unity, to the optical depth of the point where the
effective optical depth is equal to unity, i.e., 
$\tau_\nu^\ast \approx 1$ at $\tau_\nu \approx 1/\sqrt{\epsilon_\nu}$.
In view of all the approximations made in deriving Eq. (\ref{jsurf1})
(the Eddington approximation, a depth-independent $\epsilon_\nu$, etc.),
the factor $\sqrt{3}$ is inconsequential.

To proceed further, let us assume that in the deep layers we have
$4\Theta - \bar x =  2.867 \times 10^{-11} T < \bar\epsilon$, and
thus the Compton term can be neglected.  This is actually
a reasonable approximation for estimating the surface value of $\bar\nu$
for our disks, since typical local temperatures are of the order
of a few times $10^6$ K, while $\bar\epsilon \approx 10^{-4}$ in the deep
layers (note that in the deep layers $\epsilon$ is larger than near
the surface because of the higher density). Assuming also 
$\tau_{\rm tot} \gg 1$, we are left with a rough estimate
of the temperature in the deep layers, 
$T^4(\tau) \approx (3/4) T_{\rm eff}^4 \, \tau$, which follows
from equation (\ref{ener_bas}), neglecting here for simplicity a 
correction term of the order of unity coming from $\tau_w$.

We denote the temperature at which the Planck function in equation
(\ref{jsurf1}) is evaluated as $T^\ast$; it is thus given by
\begin{equation}
\label{tstar}
T^\ast = (3/4)^{1/4}\, T_{\rm eff}\, 3^{-1/8}\, \epsilon_\nu^{-1/8}\, .
\end{equation}
The dependence of $T^\ast$ on frequency arises only through the term
$\epsilon_\nu^{-1/8}$, and is therefore rather weak. We may then
replace $\epsilon_\nu^{-1/8}$ by $(\bar\epsilon)^{-1/8}$.
Similarly, if we assume that $\sqrt{\epsilon_\nu}$ varies with
frequency much more slowly than $B_\nu$ (at least around the maximum
of the Planck function that contributes most to the integrals),
then we may again use equation (\ref{nubar2}) to obtain
\begin{eqnarray}
\label{nubar}
\bar\nu = 3.83\, (kT^\ast/ h) = 3.83\, (k/h)\, (3/4)^{1/4} 3^{-1/8}\, 
(\bar\epsilon)^{-1/8} T_{\rm eff} \nonumber \\
\approx 6.5 \times 10^{10} \, (\bar\epsilon)^{-1/8}\, T_{\rm eff}\, .
\quad\quad\quad\quad \quad
\end{eqnarray}

The Compton cooling term may thus be expressed as
\begin{equation}
4\Theta - \bar x = {k \over m_{\rm e} c^2}\, (4\, T - 3.83\, T^\ast)
= 6.75\times 10^{-10}\, (T - 0.96\, T^\ast)\, ,
\end{equation}
where we left the ``formation temperature'', $T^\ast$, unspecified.
This equation is instructive because we easily see that if
$T > 0.96\, T^\ast$, Compton scattering leads to cooling of the 
surface layers,
while if $T < 0.96\, T^\ast$, Compton scattering heats the layers.
If Compton scattering is the dominant source of opacity
and emissivity, then the surface layers have the temperature
$T \approx T^\ast$.


\subsection{Surface temperature}

An estimate of the surface temperature is of considerable interest since 
it will show whether our models are in principle able to produce
high-temperature external layers, traditionally called
disk {\em coronae}.

Using the results of the previous section, we can rewrite equation
(\ref{tempg1}) at the surface (i.e. $\tau=\tau_w=0$) as
\begin{equation}
\label{tsur0}
\bar\epsilon \left({T_0 \over T_{\rm eff} } \right)^4 = {\sqrt 3\over 4} 
\left[ \bar\epsilon - {4k\over m_{\rm e} c^2} (T_0 - 0.96\, T^\ast) \right]
+ {w(0)\over \bar w}\, {1\over 4 \tau_{\rm tot}}\, .
\end{equation}
Using equation (\ref{tstar}), and introducing a normalized surface
temperature,
$t \equiv T_0/T_{\rm eff}$, we obtain the following fourth-order 
algebraic equation for $t$,
\begin{equation}
\label{tsur1}
a t^4 + b t = c\, ,
\end{equation}
where
\begin{equation}
a=\bar\epsilon\, ,\quad\quad b=2.92 \times 10^{-10} \, T_{\rm eff}\, ,
\end{equation}
and
\begin{equation}
c = 0.43\, \bar\epsilon + 2.92 \times 10^{-10} \, T_{\rm eff}\,
(\bar\epsilon)^{-1/8} + {w(0)\over \bar w}\, {1\over 4 \tau_{\rm tot}}\, .
\end{equation}
A general solution of equation (\ref{tsur1}) is possible to find
analytically, but is quite cumbersome.
However, we may consider several limiting cases:

1) $a \gg b$, i.e. $\bar\epsilon \gg 3 \times 10^{-10} \, T_{\rm eff}$\\
This is the limit of negligible Compton scattering, which applies
if either the mean photon destruction parameter $\bar\epsilon$ is
very large, or the effective temperature is very low.
In this case the solution of equation (\ref{tsur1}) is simply
$t=(c/a)^{1/4}$, i.e.
\begin{equation}
\label{tapprox1}
{T_0 \over T_{\rm eff}} \approx \left( 0.43 + 
{w(0)\over \bar w}\, {1\over 4 \tau_{\rm tot}} \,
{1\over \bar\epsilon}\right)^{1/4}\, .
\end{equation}
There are  two limiting cases in this situation. If the second term
in the brackets in negligible, then one is left with the classical
stellar-atmosphere surface temperature: $T_0 \approx 0.81\, T_{\rm eff}$.
The second term is important, if (i) $w(0)/\bar w$ is large, i.e., if
the viscosity increases towards the surface; or (ii) if the term
$\bar\epsilon\, \tau_{\rm tot}$ is small. Note that the latter term
is in fact the total Planck-mean optical depth, i.e. the total
optical depth for pure thermal absorption. 
Therefore, the surface temperature may be large (i.e., larger than
$T_{\rm eff}$)  even in the absence of Compton scattering and an
outward-increasing dissipation. This is readily explained. 
Thomson scattering does not influence the energy balance. 
Since we now have an optically thin disk with respect to the total 
Planck-mean opacity, the total energy radiated away is proportional to 
$S\, \tau$, where $S$ is the source function,
and $\tau$ the total optical depth. We take $S=B\propto T^4$, while
the total energy dissipated is proportional to $T_{\rm eff}^4$. These
energies are equal, so that we obtain $(T/T_{\rm eff})^4 \propto 1/\tau$.
In other words, an effectively thin slab must be hotter than a thicker
slab to be able to radiate the same dissipated energy.

2) $a \ll b$, i.e. $\bar\epsilon \ll 3 \times 10^{-10} \, T_{\rm eff}$.\\
This is the limit of dominant Compton scattering.
The solution of equation (\ref{tsur1}) is
$t=c/b$, i.e.
\begin{equation}
\label{tapprox2}
{T_0 \over T_{\rm eff}} \approx (\bar\epsilon)^{-1/8} +{w(0)\over \bar w}\, 
{1\over 4 \tau_{\rm tot}}{1\over 2.92 \times 10^{-10} \, T_{\rm eff}},
\end{equation}
provided that $at^4$ is also $\ll bt$, i.e., $t \ll (b/a)^{1/3}$.
When this criterion is satisfied, there are
two limiting cases. If the second term on the
right-hand side is small, then 
$T_0 \approx (\bar\epsilon)^{-1/8}\, T_{\rm eff}$.  For typical values of
$\bar\epsilon$ of the order of $10^{-6}$ to $10^{-5}$, we see that 
Compton scattering leads to a heating of the surface layers up to a factor
of 3 -- 6 above the effective temperature! 
Obviously, the surface temperature is larger for smaller
values of $\bar\epsilon$ because Compton scattering becomes more efficient
with respect to the thermal processes.
On the other hand, if the second term is important, either because
$w(0)/\bar w$ is large or the total optical depth is small,
then the solution is given by the second term of equation (\ref{tapprox2}).
When $at^4$ is not small compared to $bt$, equation (\ref{tapprox1})
is again roughly correct.

We summarize the behavior of the surface temperature in 
Figures \FIGSTA~and \FIGSTB.

\hskip -0.2in
\parbox{3.0in}{\epsfxsize=3.5in \epsfbox{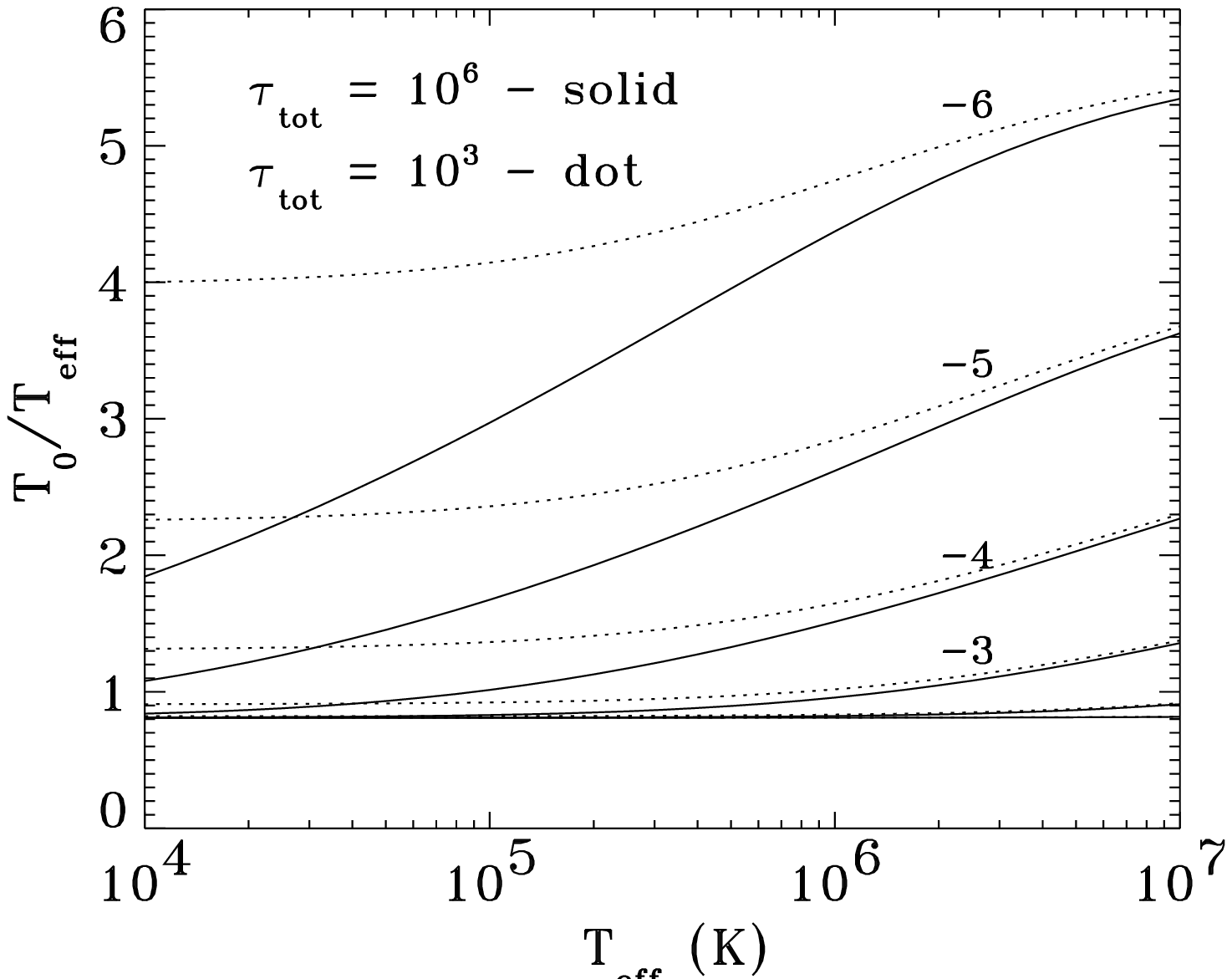}} \vskip 0.1in
\label{FIGSTA}

\centerline{\parbox{3.5in}{\small {\sc Fig.~\FIGSTA--} 
Approximate surface temperature, expressed as $T_0/T_{\rm eff}$,
displayed as a function of $T_{\rm eff}$, for different values of 
$\bar\epsilon$, and for two values of $\tau_{\rm tot}$. The individual
curves are labeled by the value of $\log \bar\epsilon$.
Solid lines are for a total optical depth $\tau_{\rm tot} = 10^6$, and
dotted lines are for $\tau_{\rm tot} = 10^2$.
}}

\vskip0.1in
\addtocounter{figure}{1}

For simplicity, we assume constant viscosity, i.e. $w(0)/\bar w=1$;
we will study the effects of an increasing viscous dissipation towards the
surface in a subsequent paper.
Figure~\FIGSTA displays the surface temperature, expressed as $T_0/T_{\rm eff}$
as a function of $T_{\rm eff}$, for different values of $\bar\epsilon$,
and for two values of $\tau_{\rm tot}$.  Figure~\FIGSTB displays the surface
temperature as a function of the total optical depth, for several values
of $\bar\epsilon$ and $T_{\rm eff}$. 

\hskip -0.2in
\parbox{3.0in}{\epsfxsize=3.5in \epsfbox{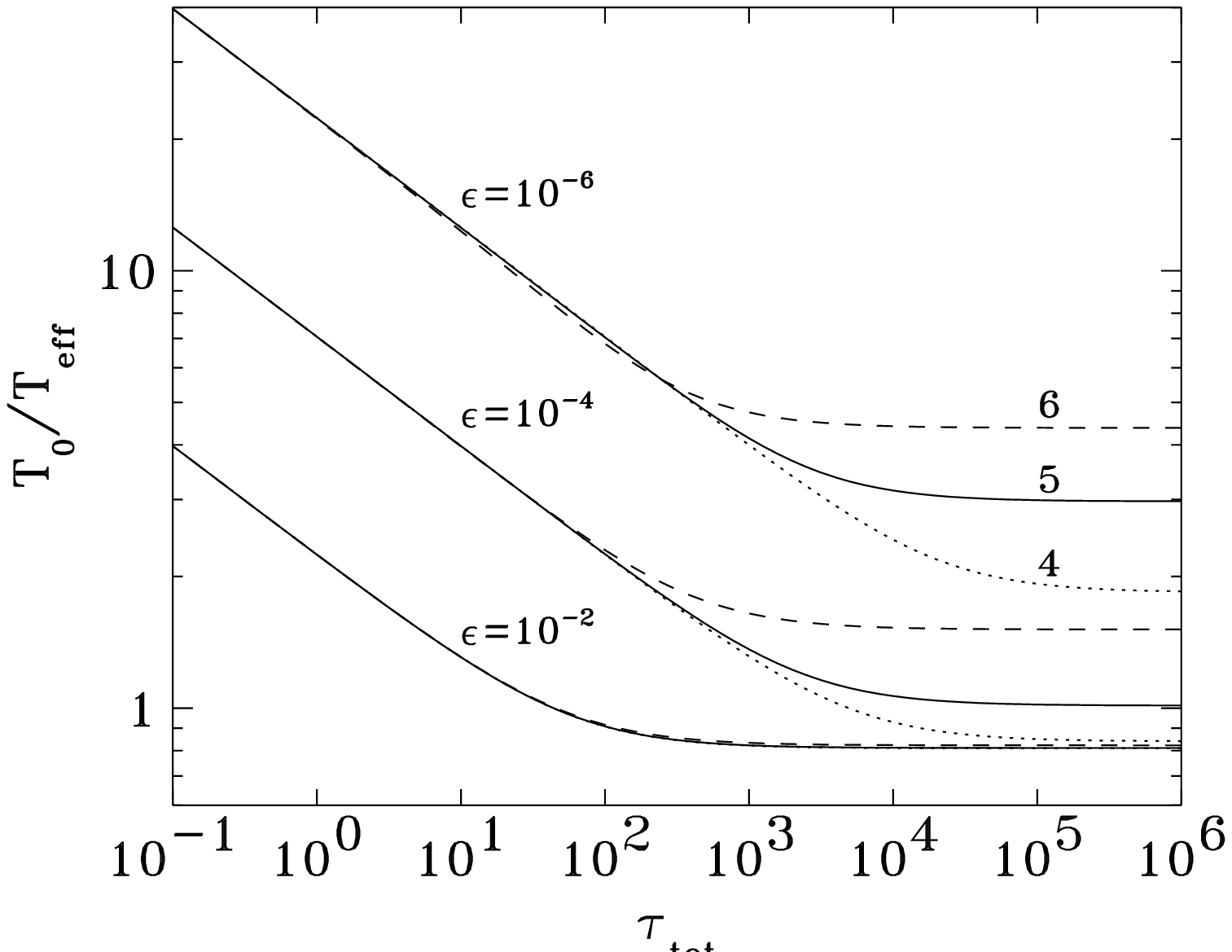}} \vskip 0.1in
\label{FIGSTB}

\centerline{\parbox{3.5in}{\small {\sc Fig.~\FIGSTB --} 
Approximate surface temperature, expressed as $T_0/T_{\rm eff}$,
displayed as a function of $\tau_{\rm tot}$, for different values of 
$\bar\epsilon$, and for three values of $T_{\rm eff}$. The individual
curves are labeled by the value of $\bar\epsilon$ and by the value
of $\log T_{\rm eff}$.
Solid lines have $T_{\rm eff} = 10^5$ K, dotted lines have
$T_{\rm eff} = 10^4$ K, and dashed lines have $T_{\rm eff} = 10^6$ K.
}}

\vskip0.1in
\addtocounter{figure}{1}

We now turn to the accurate numerical models, and show that
the above estimates agree reasonably well with exact numerical results.


\section{Results}


\subsection{Comparison to the models from Paper III}

We shall first present a comparison of models computed with and
without Comptonization.  We take the hottest disk from our grid
of Paper~III, where the effects of Comptonization are
expected to be the largest. This is a disk around a black hole with
$M=(1/8)\times10^9\ $M$_\odot$ and mass accretion rate 
$\dot M = 1/4\ $M$_\odot$/yr (i.e. with luminosity
$L \approx 0.3\, {\rm L}_{\rm Edd}$).  To isolate the effects of
Comptonization, we assume, in agreement with Paper~III, that the disk
consists of hydrogen and helium only. Also following Paper~III, we take the
local viscosity
to be described by a two-step power law, with a constant kinematic viscosity
in the inner 99\% of the disk mass, while the viscosity is assumed to decrease
as $m^{-2/3}$ in the outermost 1\% of the mass (see Paper~III for 
details).
The reason we chose an artificial decrease of viscosity with
height was to prevent a ``thermal catastrophe" of the disk in the
low optical depth regions (e.g., Shaviv \& Wehrse 1986).
In the case of H-He models without Comptonization, there are no
efficient mechanisms that are able to radiate the dissipated
energy away when the atmosphere becomes optically thin in the
most opaque transition; the atmosphere thus undergoes a thermal
runaway.  This is obviously prevented when one considers Compton
instead of Thomson electron scattering, as we have shown 
analytically in the previous section, and will demonstrate
numerically in this and subsequent sections.  We show the
models with vertically decreasing dissipation only for the purpose of
direct comparison of the present results to those of Paper~III.
%
%

%
\hskip -0.2in
\parbox{3.0in}{\epsfxsize=3.5in \epsfbox{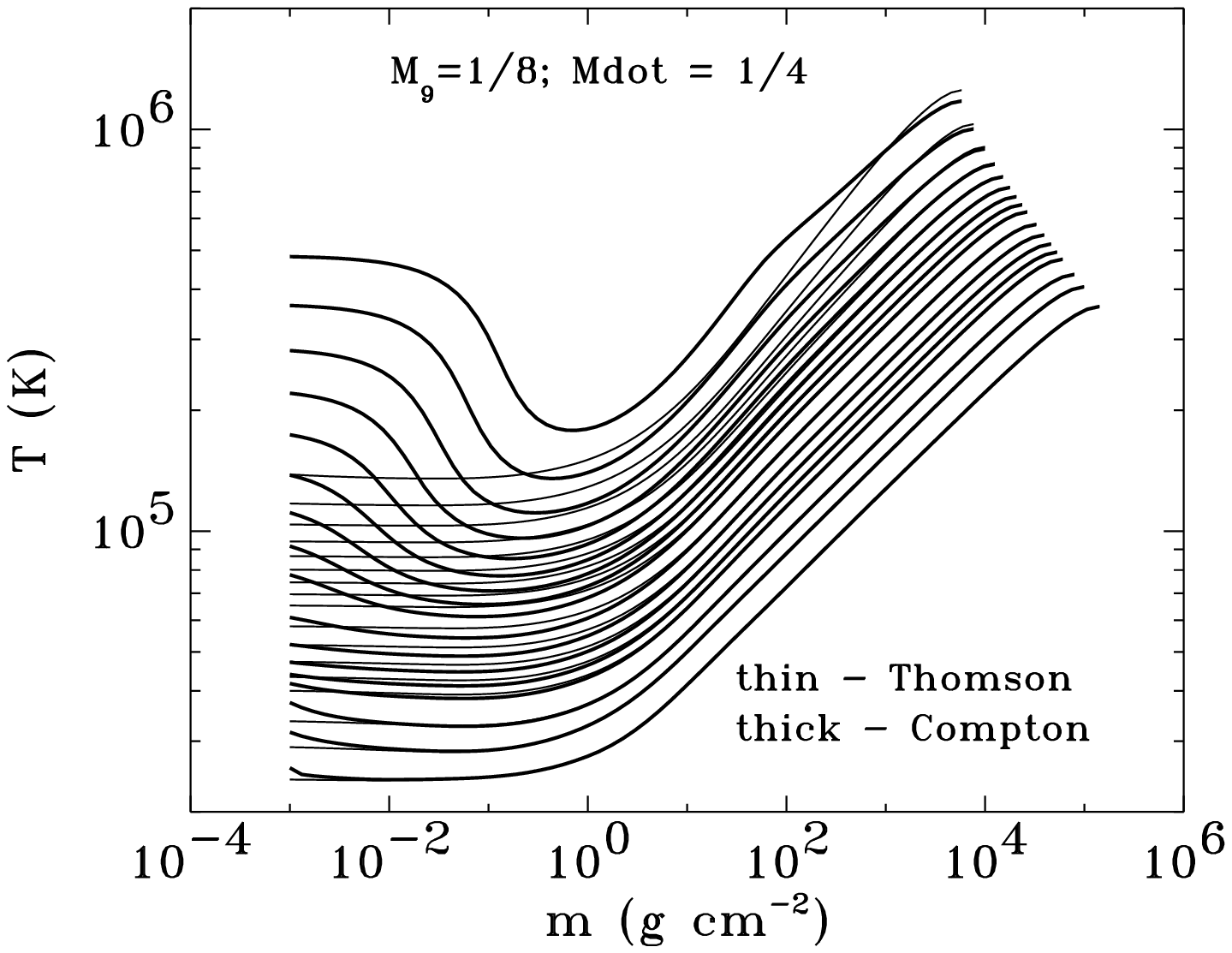}} \vskip 0.1in
\label{FIGTA}

\centerline{\parbox{3.5in}{\small {\sc Fig.~\FIGTA --} 
Temperature as a function of depth for selected annuli. 
From top to bottom, the curves correspond to radii $R/R_g = $
2, 3, 4, 5, 6, 7, 8, 9, 10, 12, 14, 16, 18, 20, 25, 30,
40, 50, 60, 70, 80, and 90.
Thin lines represent the models from Paper~III computed without Comptonization,
while thick lines correspond to models with Comptonization.
}}

\vskip0.1in
\addtocounter{figure}{1}

Figure~\FIGTA~displays the local electron temperature as a function
of position for selected annuli. The position is expressed as 
column mass, $m$, above a given depth. 
The models with Comptonization exhibit a temperature rise at the surface,
even for the case of viscosity decreasing outward.  This is easily
explained by the considerations presented in the previous section.
The hottest annuli have $T_{\rm eff} \approx 2\times 10^5$ K. 
The typical value of $\bar\epsilon$ is about $10^{-5}$, so that
$\epsilon < 3 \times 10^{-10} \, T_{\rm eff} \approx 6 \times 10^{-5}$,
and thus we are in the regime of important Compton scattering, where
Eq. (\ref{tapprox2}) applies. Since viscosity is forced to decrease 
outward, the second term on the right-hand side of Eq. (\ref{tapprox2})
is negligible, and we therefore find
$T/T_{\rm eff}\approx2$, which is indeed
seen in Figure~\FIGTA.  The cooler annuli have progressively larger values of
$\bar\epsilon$ and lower $T_{\rm eff}$, and therefore are less and 
less influenced by Comptonization.
Again, this is nicely verified by numerical calculations.

%

%
\hskip -0.2in
\parbox{3.0in}{\epsfxsize=3.5in \epsfbox{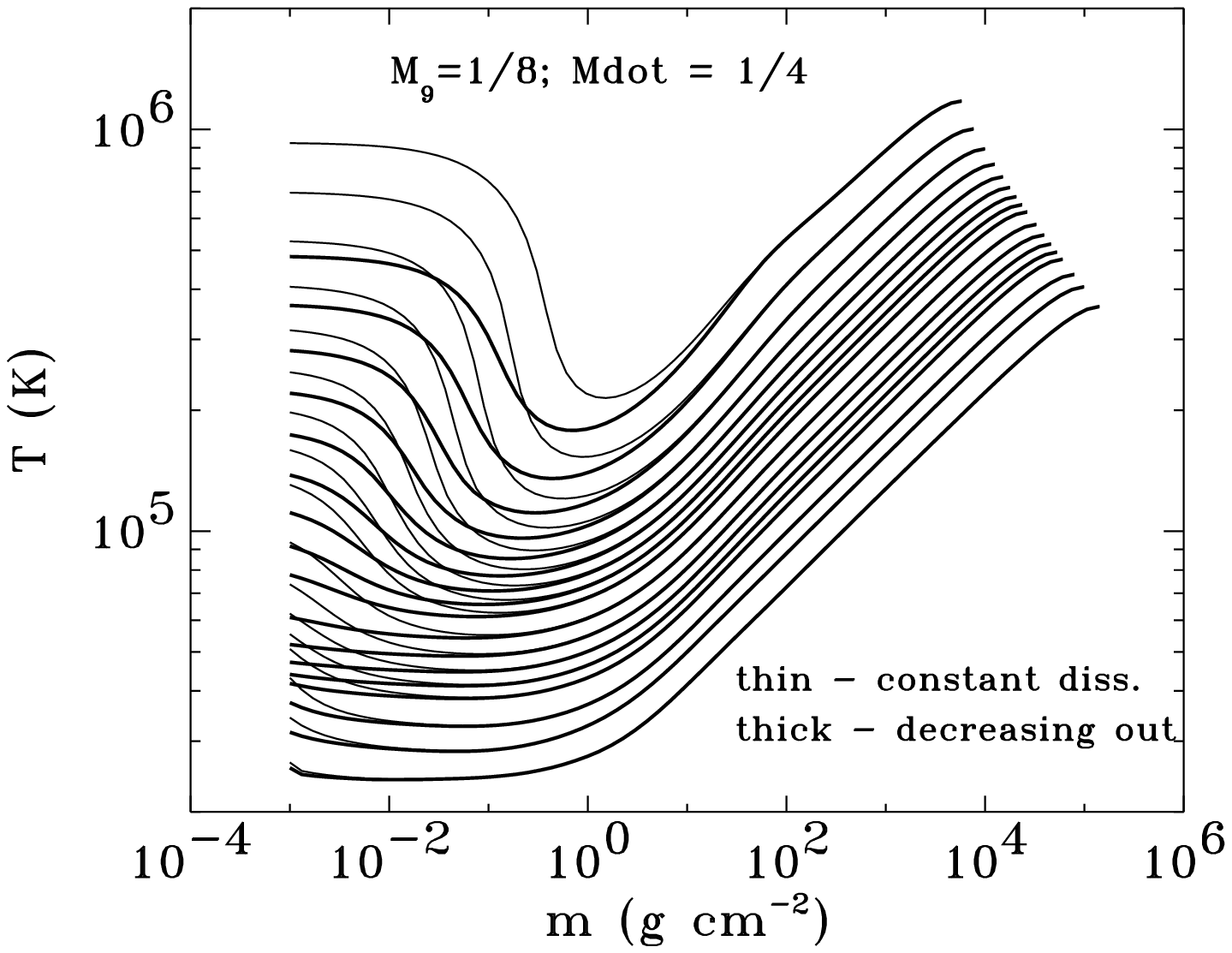}} \vskip 0.1in
\label{FIGTB}

\centerline{\parbox{3.5in}{\small {\sc Fig.~\FIGTB --} 
Temperature as a function of depth for selected individual annuli. 
The curves correspond to radii $R/R_g$ (from top to bottom)
2, 3, 4, 5, 6, 7, 8, 9, 10, 12, 14, 16, 18, 20, 25, 30,
40, 50, 60, 70, 80, and 90.
Thin lines represent the models with constant dissipation, while
thick lines represent the models with decreasing dissipation towards the
surface.
}}

\vskip0.1in
\addtocounter{figure}{1}

To see the effect of changing the dissipation law on the temperature structure,
we have computed the same models as above, but keeping the kinematic viscosity
constant with depth, i.e. without an artificial decrease of viscosity
at the surface. The temperature structure for the individual annuli
is displayed in Figure~\FIGTB. Note that models with constant
dissipation were impossible to construct with the original assumption
of Thomson scattering, because the temperature exhibited a
runaway instability behavior.
The behavior of the surface temperature is again easily explained using
Eq. (\ref{tapprox2}).  In this case, we have $w(0)/\bar w = 1$.
Moreover, 
for the hottest annuli we have $T_{\rm eff} \approx 2\times 10^5$ K,
and $\tau_{\rm tot} \approx 2 \times 10^3$,
so the second term in equation (\ref{tapprox2}) is about 2.
Since the first
term is also roughly equal to 2, we see that in this particular
case the Compton heating and the viscous heating have a comparable
effect on the surface temperature.
The resulting $T/T_{\rm eff}$ should be around 4, which is again nicely 
verified by numerical calculations.

%

%
\hskip -0.2in
\parbox{3.0in}{\epsfxsize=3.5in \epsfbox{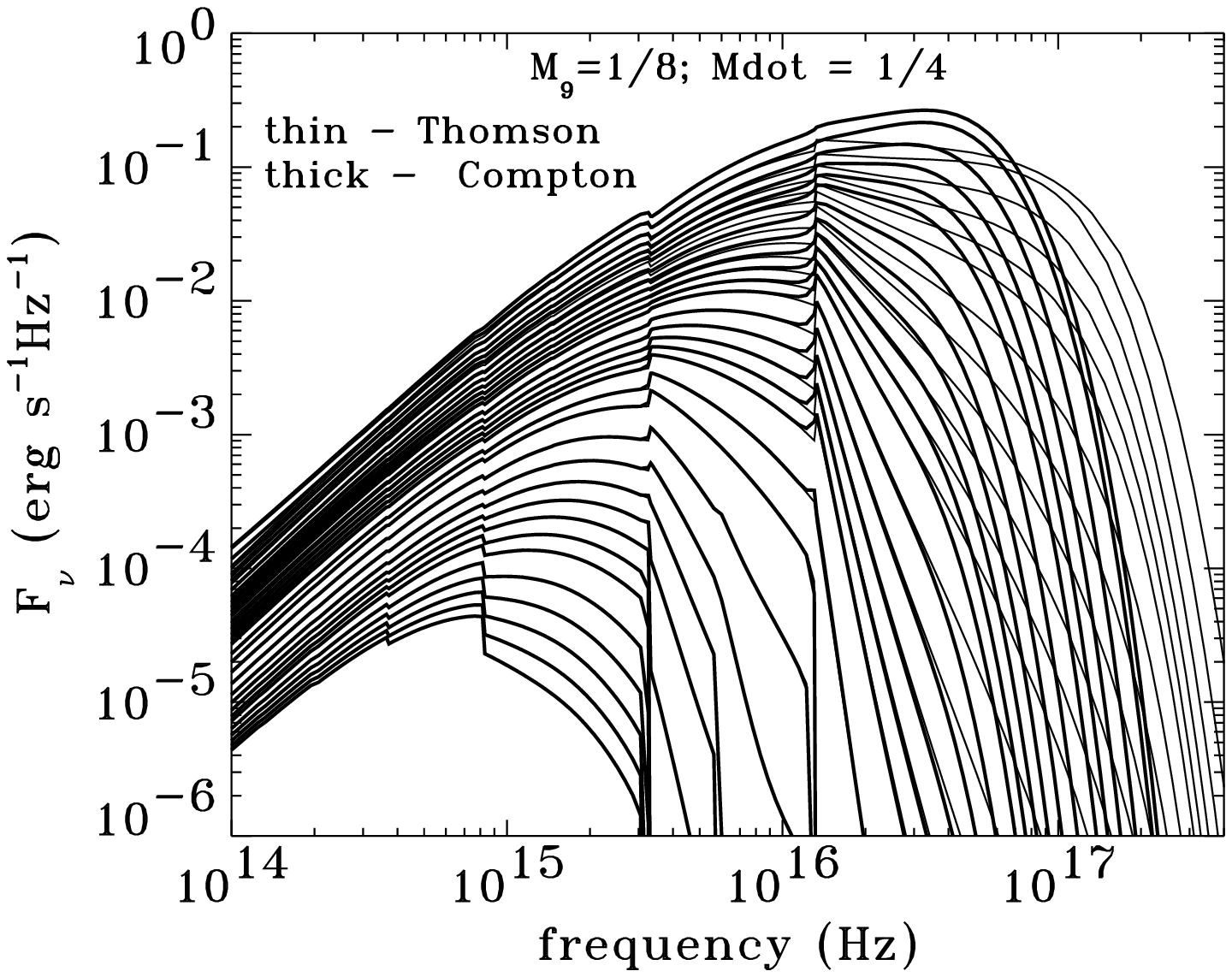}} \vskip 0.1in
\label{FIGFLA}

\centerline{\parbox{3.5in}{\small {\sc Fig.~\FIGFLA --} 
Local emergent flux for the individual annuli. 
From top to bottom, the curves correspond to radii $R/R_g =$
1.5, 2, 2.5, 3, 3.5, 4, 5, 6, 7, 8, 9, 10, 12, 14, 16, 18, 20, 25, 30,
40, 50, 60, 70, 80, 90, 100, 120, 140, 160, 180, and 200.
Thin lines represent the models from Paper~III computed without Comptonization,
and thick lines represent the models with Comptonization. 
}}

\vskip0.1in
\addtocounter{figure}{1}

From a practical point of view, the most important feature is
the behavior of the emergent flux from the individual annuli when
Comptonization is taken into account. Figure~\FIGFLA~displays
a comparison between the model without (thin lines) and with (heavy
lines) Comptonization for the individual annuli displayed in 
previous figures. 

The models with constant and with decreasing
kinematic viscosity at the surface are indistinguishable in the plot.
This is because only the surface layers are influenced by changes
in the dissipation law, while the emergent continuum radiation is formed
in much deeper layers. The emergent flux would be influenced if the
dissipation increases outward, in which case a hot corona (with
temperatures of the order $10^8$ to $10^9$ K) is formed. Such a corona would
produce a significantly increased X-ray flux.  
We will present a more extended study 
of this phenomenon in the next paper of this series.

Comptonization in hot annuli leads to the well-known increase
of the flux in the EUV and the soft X-ray region ($\nu < 10^{17}$ s$^{-1}$).
For larger frequencies, the flux drops fast, so that
the flux for the Comptonized model is lower than the flux
for the original, Thomson-scattering, model.  
An analogous behavior was also found for the emergent spectra
of (hotter and denser) neutron star atmospheres (Madej 1991).
This behavior is readily
explained by examining the net photon energy gain (or loss) as a function
of frequency. As discussed in Section 3 (see also Rybicki \&~Lightman
1979), the net energy gain of a photon with frequency $\nu$ is given by
$4 \Theta - x$.  Since we are dealing with a depth-dependent situation, one
has to evaluate the appropriate $\Theta$ at the effective depth
of formation, 
$\tau_\nu \approx \tau_\nu^{\rm form} \approx 1/\sqrt{\epsilon_\nu}$.
Note that when $\tau_\nu^{\rm form} > \tau_{\rm tot}$, where
$\tau_{\rm tot}$ is the optical depth at the midplane (generally, it is
a frequency-dependent quantity, but the dependence is very
weak in the case of dominant electron scattering), the disk is 
{\em effectively thin}.  The effective depth of formation should
thus be more generally given by 
$\tau_\nu^{\rm form} \approx \min(1/\sqrt{\epsilon_\nu}, \tau_{\rm tot})$.
Nevertheless, we will assume an effectively thick disk in the following
discussion.
The temperature at the formation depth is thus given by
[see also Eq. (\ref{tstar})], 
$T^\ast_\nu \approx T_{\rm eff} \, \epsilon_\nu^{-1/8}$.

%

%
\hskip -0.2in
\parbox{3.0in}{\epsfxsize=3.5in \epsfbox{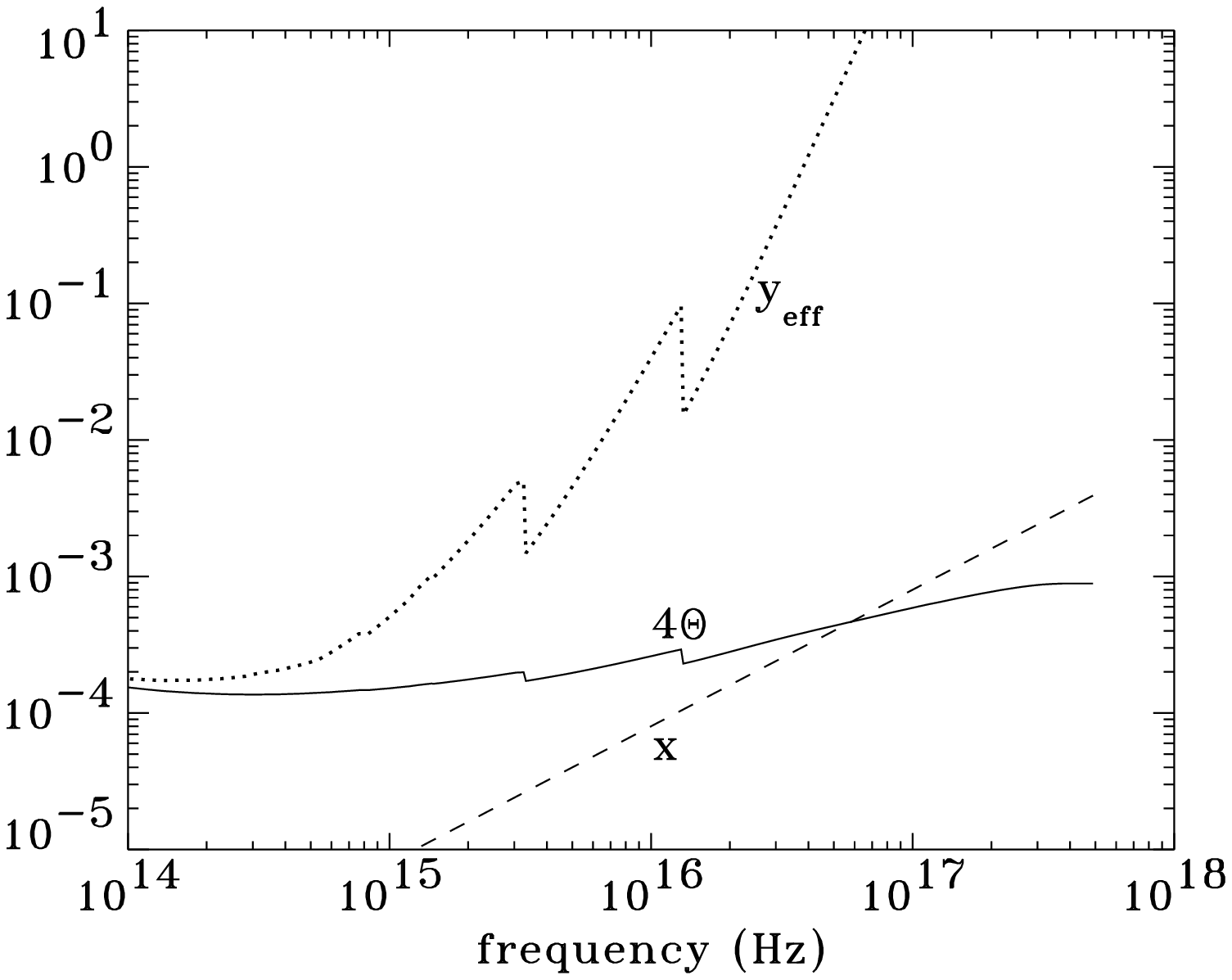}} \vskip 0.1in
\label{FIGC}

\centerline{\parbox{3.5in}{\small {\sc Fig.~\FIGC --} 
Computed value of $4 \Theta(T^\ast_\nu)$ compared
to $x$ as a function of frequency. The Compton parameter $y_{\rm eff}$
is also displayed.  All the quantities are dimensionless numbers.
}}

\vskip0.1in
\addtocounter{figure}{1}

In Figure~\FIGC~we display the computed value of $4 \Theta(T^\ast_\nu)$ compared
to $x$ as a function of frequency.  We see that for 
$\nu \lta 6 \times 10^{16}$, the term $4\Theta-x$ is positive,
and therefore Comptonization leads to a net {\em increase} in
photon frequency, while for $\nu \gta 6 \times 10^{16}$ the
photon frequency is {\em decreased} by Comptonization.  These values tell us
how the photon frequency is shifted, but do not tell us about
the number of available scatterings and thus about the resulting
effects of the photon frequency shifts on the emergent spectrum.
In order to verify that Comptonization in fact has a significant effect on the
spectrum, we also show the effective Compton parameter $y_{\rm eff}$, which is
defined by (Rybicki \&~Lightman 1979)
\begin{equation}
y_{\rm eff} = 4 \Theta \max[\tau_\nu^{\rm form}, (\tau_\nu^{\rm form})^2] \, .
\end{equation}
Here the factor $\max[\tau_{\rm form}, (\tau_{\rm form})^2]$ expresses
the total number of scatterings of a photon with frequency $\nu$.
Since in our case $\tau_\nu^{\rm form} \approx 1/\sqrt{\epsilon_\nu}$
and thus $\tau_\nu^{\rm form} \gg 1$, we have
\begin{equation}
\label{eqyeff}
y_{\rm eff} \approx 4 \times 10^{-10} \, T_{\rm eff}\, \epsilon_\nu^{-9/8} \, .
\end{equation}
We can roughly estimate $\epsilon_\nu$ and therefore $y_{\rm eff}$
as follows.  Let us assume for simplicity a relatively hot plasma where
all hydrogen and helium are almost completely ionized, and that the only
sources of thermal opacity are the Lyman continua and the free-free transitions 
of hydrogen and ionized helium. The bound-free and free-free opacities are 
thus given by (see, e.g., Mihalas 1978)
\begin{eqnarray}
\kappa_{\rm bf}(\nu) \approx 1.17\times 10^{14} \, T^{-3/2}\, 
n_{\rm e}\, \nu^{-3}\,\quad\quad\quad\quad \nonumber \\
\left[ n_{\rm p}\, e^{0.158/T_6}\, H(\nu, \nu_{\rm H})\, + 
Z^4\, n_{\rm He III}\, e^{0.631/T_6}\, H(\nu, Z^2 \nu_{\rm H})\right]\, ,
\end{eqnarray}
and
\begin{equation}
\kappa_{\rm ff}(\nu) \approx 3.69\times 10^{8}\, T^{-1/2}\, n_{\rm e}\, 
\nu^{-3}\, [n_{\rm p}  + Z^2\, n_{\rm He III}]\, ,
\end{equation}
where $Z=2$ is the effective charge of the ionized helium; $n_{\rm p}$
and $n_{\rm He III}$ are the number density of protons and of 
helium nuclei, respectively; and $T_6\equiv T/10^6$.  Here we assume LTE
for the hydrogen and ionized helium ground-state populations.
We also assume that the bound-free and free-free Gaunt factors 
are equal to unity. 
$H(\nu, \nu_0)$ is the unit step function, defined in such
a way that $H=0$ for $\nu < \nu_0$, and $H=1$ for $\nu \geq \nu_0$;
$\nu_{\rm H} = 3.29\times 10^{15}$ s$^{-1}$ is the frequency 
of the hydrogen Lyman edge.
Since we assume almost completely ionized plasma, then from the
charge conservation equation we obtain $n_{\rm p} = n_{\rm e}/(1+2Y)$
and $n_{\rm He III} = n_{\rm e} Y/(1+2Y)$.  Substituting these values, 
and assuming the solar abundance of helium, $Y=0.1$,
we obtain for  $\epsilon_\nu$ 
\begin{equation}
\label{appeps1}
\epsilon_\nu \approx {\kappa_{\rm bf}(\nu) +\kappa_{\rm ff}(\nu) \over
n_{\rm e} \sigma_{\rm T}} \approx
6.5 \times 10^{-5}\, n_{14}\,  T_6^{-1/2}\, \nu_{16}^{-3}\, \gamma_\nu(T)\, ,
\end{equation}
where $n_{14} = n_{\rm e}/10^{14}$, $\nu_{16} = \nu/10^{16}$,
and $\gamma$ is a function of order of unity that depends only weakly
on $T$ and $\nu$ (at least in the temperature range $T \lta 2 \times 10^5$ K
relevant here).
\begin{equation}
\gamma_\nu(T) = 1 + {0.226\over T_6} \left[ 
e^{0.158/T_6}\, H(\nu, \nu_{\rm H}) + 1.6\, e^{0.631/T_6}\,
H(\nu, 4 \nu_{\rm H}) \right]\, .
\end{equation}
Equation (\ref{appeps1}) is valid for any temperature and electron density.
We may also specify the corresponding $\epsilon$ at the effective depth of
formation, $\epsilon_\nu^\ast \equiv \epsilon_\nu(T_\nu^\ast)$, where
$T_\nu^\ast$ is given by 
$T^\ast_\nu \approx T_{\rm eff} \, (\epsilon_\nu^\ast)^{-1/8}$.
Substituting for $\epsilon$ from equation (\ref{appeps1}),
we obtain
\begin{equation}
T^\ast_{\nu, 6} \approx 3.6\, T_{{\rm eff}, 6}^{16/15}\, n_{14}^{-2/15}\,
\nu_{16}^{2/5}\, ,
\end{equation}
[notice that the scaling $\Theta(T^\ast_\nu) \propto \nu^{2/5}$
is nicely seen in Figure~\FIGC],
and thus
\begin{equation}
\epsilon_\nu^\ast \approx 3.2\times 10^{-5}\, T_{{\rm eff}, 6}^{-8/15}\, 
n_{14}^{16/15}\, \nu_{16}^{-16/5}\, .
\end{equation}
Here we use the same notations as before, namely
$T_{{\rm eff}, 6} \equiv T_{\rm eff}/10^6$, and 
$T^\ast_{\nu, 6} \equiv T^\ast_{\nu}/10^6$.
Finally, from equation (\ref{eqyeff}) we have
\begin{equation}
\label{yeff2}
y_{\rm eff} \approx 44\, T_{{\rm eff}, 6}^{24/15}\, 
n_{14}^{-18/15}\, \nu_{16}^{18/5}\, .
\end{equation}
Again, the approximate scaling $y_{\rm eff} \propto \nu^{18/5}$ is verified
in Figure~\FIGC.

We see from Figure~\FIGC~that $y_{\rm eff} \gta 1$ 
for $\nu \gta 3 \times 10^{16}$, and thus Comptonization
should significantly influence the emergent spectral energy distribution for
these frequencies. 
From Figure~\FIGC~we deduce that the emergent flux for frequencies
$3 \times 10^{16} \lta \nu \lta 6 \times 10^{16}$ should be enhanced over
the model with Thomson scattering only, while the opposite should
apply for higher frequencies, which is indeed demonstrated
by the exact results shown in Figure~\FIGFLA.  In fact, the enhancement of
flux caused by Compton scattering is seen already for lower frequencies,
which is a consequence
of the higher temperature for the Comptonized model (cf. Figure~\FIGTA).

An inspection of Figure~\FIGFLA~shows that the emergent flux is 
influenced only for the hottest annuli, those with effective temperature 
above 150,000 K; the cooler annuli are influenced only marginally. 
This is easily seen from equation (\ref{yeff2}). The effective temperature
varies with radial coordinate roughly as $T_{\rm eff} \propto R^{-3/4}$,
so that $T_{\rm eff}^{24/15} \propto R^{-6/5}$. Estimating the
appropriate electron density is much more complex. We consider
two limits. In completely radiation-pressure dominated disks, and
neglecting relativistic corrections, we may consider the mass density
at the effective depth of formation to be equal to the vertically averaged
mass density, $\rho_0$. Thus $n_{\rm e} \propto \rho_0 \propto R^{3/2}$.
In this limit, $y_{\rm eff}\propto R^{-3}$, i.e., it decreases sharply
with increasing radial distance.  The opposite limit is to assume 
the electron density roughly constant or only slightly increasing
with $R$ (see, e.g. Figure~4 of Paper~III), which follows from an 
interplay between the higher midplane density and a sharper  decrease
of vertical density profile for increasing radial coordinate. Even in
this extreme case, $y_{\rm eff}$ decreases faster than linearly with
increasing $R$.  These considerations explain why the Comptonization
influences mostly the inner annuli, while the outer annuli are influenced
progressively less and less.

To see the effects of Comptonization on the hydrogen and He~II Lyman edges 
more clearly, we present in Figure~{\FIGFLB~a blow-up of 
Figure~\FIGFLA.
It is clear Comptonization indeed considerably smears the He~II Lyman
edge. The hydrogen Lyman edge is smeared as well, although the
effect is much smaller because Comptonization is much less
efficient at the relatively low frequencies corresponding to the
hydrogen Lyman edge. 

%

%
\hskip -0.2in
\parbox{3.0in}{\epsfxsize=3.5in \epsfbox{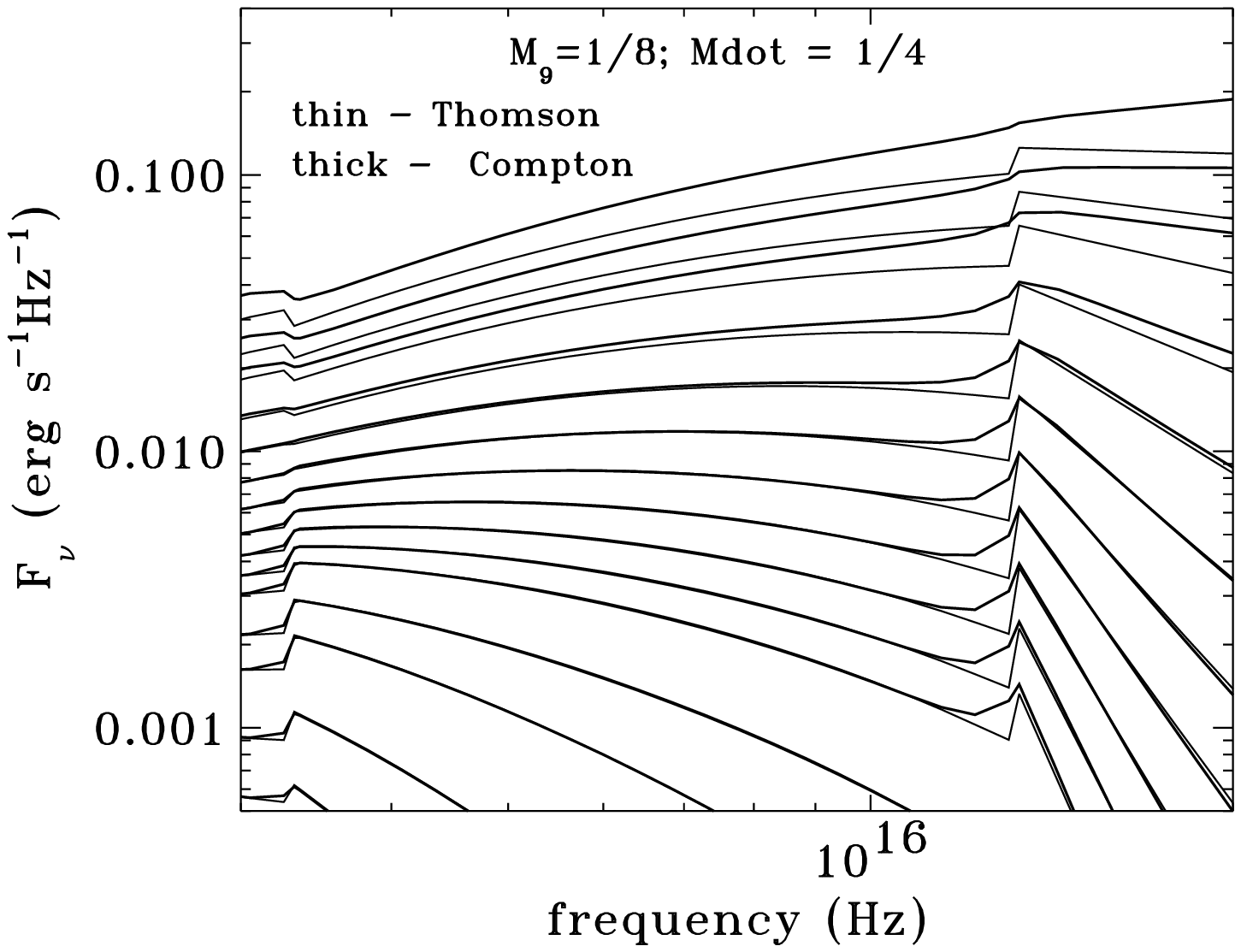}} \vskip 0.1in
\label{FIGFLB}

\centerline{\parbox{3.5in}{\small {\sc Fig.~\FIGFLB --} 
A blow-up of Figure~\FIGFLA showing the region of the hydrogen
and He~II Lyman edges. 
From top to bottom, the curves correspond to local emergent flux 
for the individual annuli at radii $R/R_g =$
2,  3, 4, 5, 6, 7, 8, 9, 10, 12, 14, 16, 18, 20, 25, and 30
Thin lines represent the models from Paper~III computed without Comptonization,
and thick lines represent the models with Comptonization. 
}}

\vskip0.1in
\addtocounter{figure}{1}

Finally, Figure~\FIGINT~presents disk-integrated spectra
for the above disk model. As in Paper~III,
the spectrum is found by integrating the total emergent intensity
over the disk surface using our relativistic transfer function code (Agol
1997).  The transfer function computes the trajectories of photons
from infinity to the disk plane, finding the radius, redshift,
and intensity at each angle at infinity for a given observation angle
(Cunningham 1975).  We neglect the effects of radiation that returns to the
accretion disk.

%

%
\hskip -0.2in
\parbox{3.0in}{\epsfxsize=3.5in \epsfbox{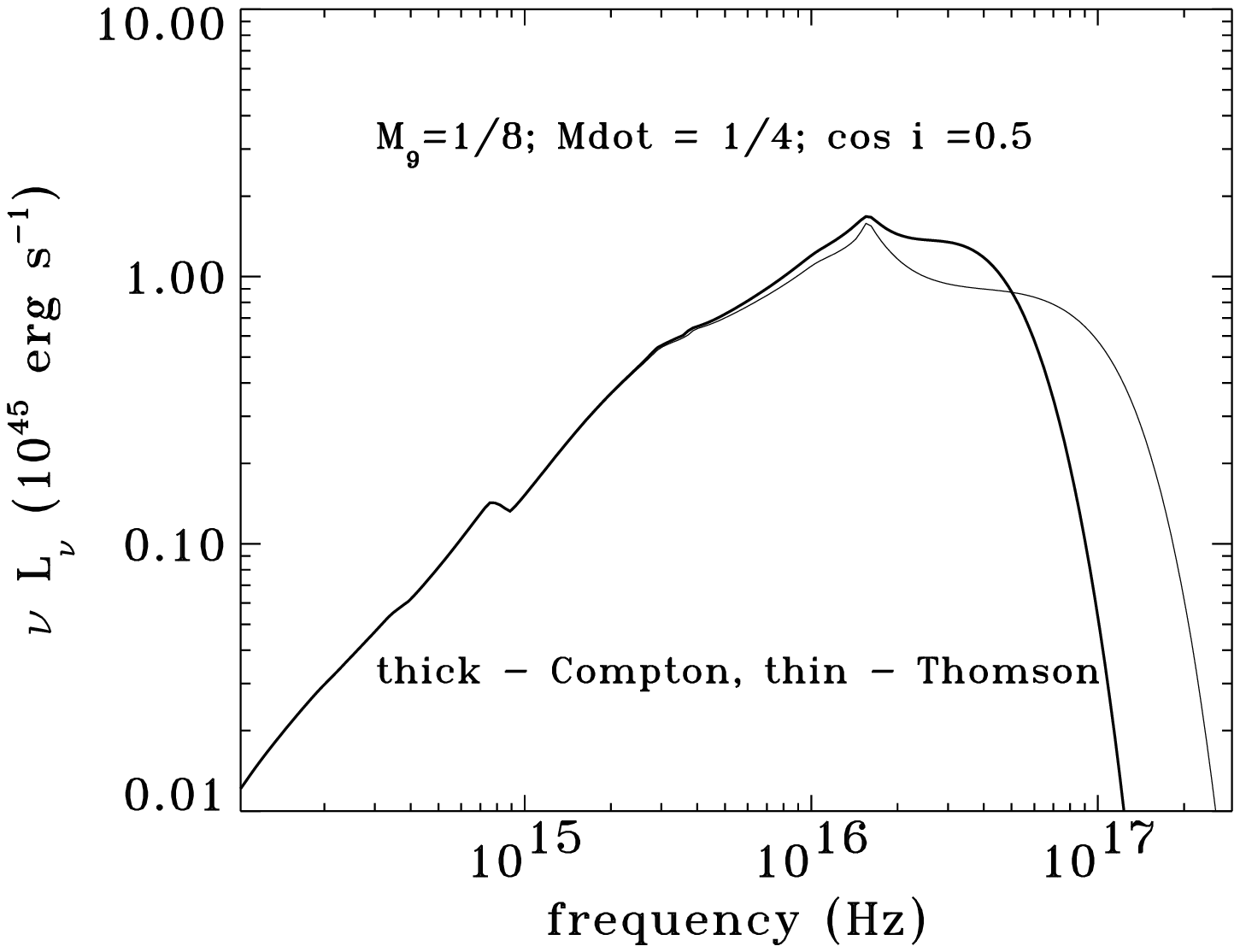}} \vskip 0.1in
\label{FIGINT}

\centerline{\parbox{3.5in}{\small {\sc Fig.~\FIGINT --} 
Integrated spectral energy distribution $\nu L_\nu$ 
(in erg s$^{-1}$) for a disk with $M=(1/8) \times 10^9\ $M$_\odot$,
$\dot M = 1/4 $M$_\odot$ yr$^{-1}$, (i.e. $L/{\rm L}_{\rm Edd} = 0.3$), and
$\cos i = 0.5$, for the model with (bold line) and without (thin line)
Comptonization.
}}

\vskip0.1in
\addtocounter{figure}{1}

Comptonization hardly affects the optical and ultraviolet portions of
the integrated spectrum, but it has a more dramatic effect at higher energies.
The flux distribution around the He~II Lyman limit is generally smoother;
the flux in the region between $2 \times 10^{16}$ and $5 \times 10^{16}$ Hz
(i.e., roughly between 0.1 and 0.2 keV) is about 30\% larger
if Comptonization is taken into account. About 10 percent of the total luminosity
is shifted from the high energy tail (above $6\times10^{16}$ Hz) to the
peak of the spectrum (between 1.5 and $4\times10^{16}$ Hz.)

We stress that, because they had larger central masses and therefore
lower central temperatures, the effects of Comptonization for all other disks
presented in Paper~III should be smaller.  Therefore, we conclude that
neglecting Comptonization made little difference to the grid of models
presented in that paper.
This conclusion is in general agreement with the results of Laor \&
Netzer (1989), who found Comptonization to be unimportant for
disks with $T_{\rm eff} \approx 10^5$ K.


\subsection{Comparison with Previous studies}

Studies of the influence of Comptonization on disk atmospheres were
pioneered by Ross, Fabian, \&~Mineshige (1992 - hereafter referred to as RFM).
We now present a comparison with the models of RFM, 
who computed models of AGN accretion
disks taking into account the effects of Compton scattering and
some NLTE effects. Specifically, they assumed a constant
(depth-independent) density in the $z$-direction, and solved
the energy balance equation, the radiative transfer equation (but in
the diffusion approximation),
and the rate equations for the first three levels of H~I and He~II.
In the latter equations, the radiative rates in the bound-bound
transitions were treated by means of escape probability theory.

In order to compare our treatment of Comptonization to theirs,
we have computed a set of test models as similar to RFM as possible. 
We have kept density
fixed, and consider only three levels for H~I and He~II. Otherwise
we have used our procedure to solve all the structural equations
except hydrostatic equilibrium. Since our program does not
use escape probabilities, we have determined the bound-bound
radiative rates exactly, by solving the transfer equation with
15 frequency points per line.

We compare a single annulus at $R/R_g=9.8$, taken from
a disk with $M=10^7$~M$_\odot$ (Schwarzschild black hole)
and $L/L_{\rm edd}=0.3$ (i.e.
$\dot M = 0.1$~M$_\odot/{\rm yr}$) to the
corresponding RFM model, kindly supplied to us by Randy Ross --
see Figure~\FIGR.  Results for other annuli are analogous.
The agreement of the computed flux
in the continuum is very good, although there are some differences,
particularly at the shortest and the longest wavelengths. 
The difference in the region of the flux peak is about 10\%,
which is likely explained by the RFM treatment of the 
inner boundary condition for the radiative transfer equation.
For the frequencies at which the total effective optical
thickness at the disk midplane is larger than unity, RFM assume
that the radiation at the effective optical depth
equal to unity is in LTE at the temperature given
by an approximate expression [their Eqs. (15) and (16)].
In the present case, the effective optical thickness is slightly
above unity for $\nu < 10^{17}$, and reaches values of about 10
at the He~II Lyman discontinuity. Therefore the RFM treatment
of the inner boundary condition is somewhat inaccurate for
these frequencies.

The predicted profiles for the hydrogen and helium lines differ
significantly, and some are even flipped from emission to absorption.
This is very likely explained by differences
in treating the lines, in particular the escape probability
treatment adopted by RFM.  Another reason may be a rather
coarse wavelength resolution, and consequently low number
of frequency points, adopted by RFM. 
As mentioned earlier, we will perform a detailed
study of the effects of spectral lines on the vertical structure and
emergent spectra for AGN disks in a future paper of this
series.

%

%
\hskip -0.2in
\parbox{3.0in}{\epsfxsize=3.5in \epsfbox{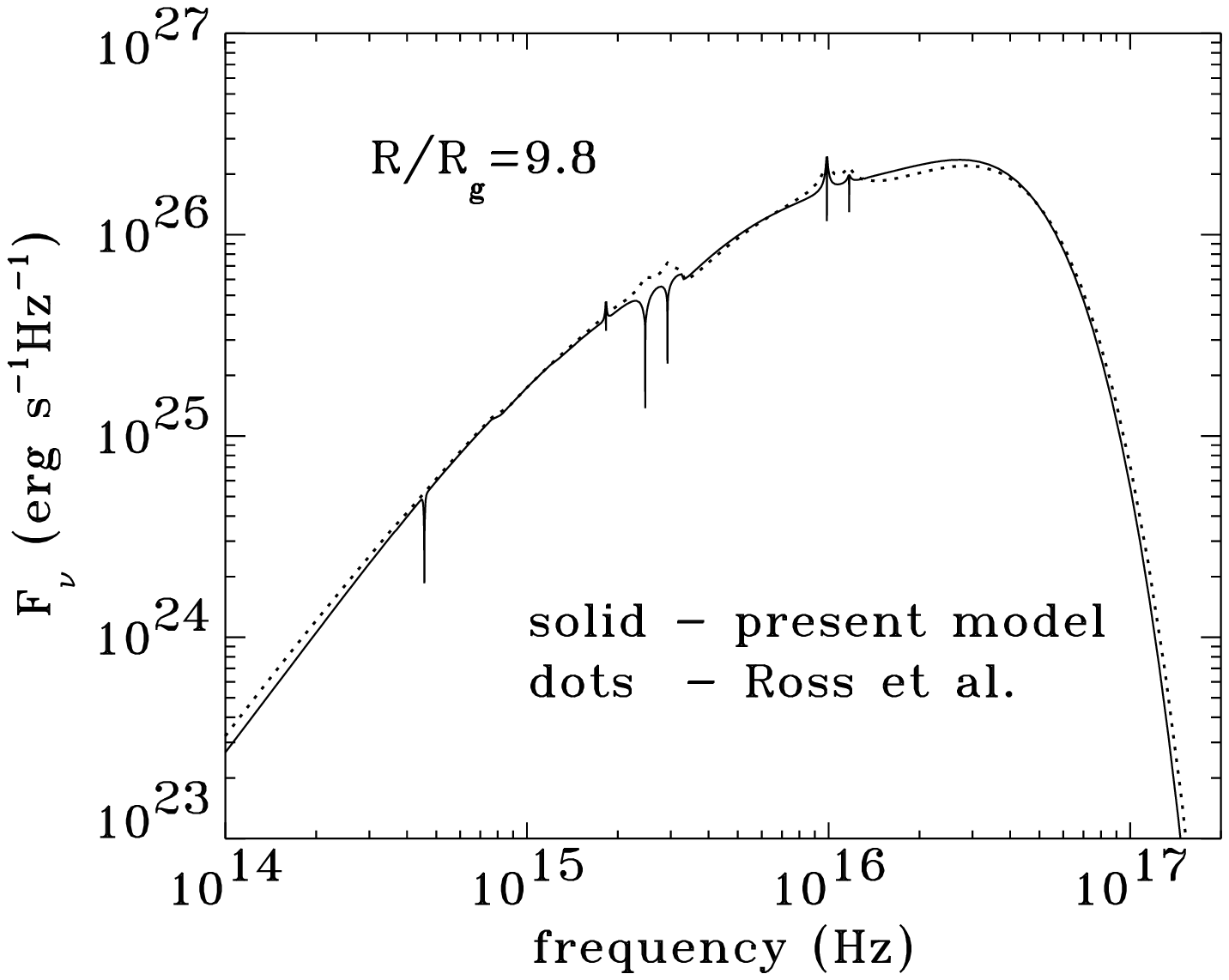}} \vskip 0.1in
\label{FIGR}

\centerline{\parbox{3.5in}{\small {\sc Fig.~\FIGR --} 
A comparison of a sample model for one annulus at $R/R_g=9.8$,
for a disk with $M=10^7~$M$_\odot$ and $L/L_{\rm edd}=0.3$ (i.e.
$\dot M = 0.1$~M$_\odot/{\rm yr}$) computed by TLUSDISK (solid line) with the
corresponding Ross, Fabian, Mineshige (1992) model (dotted line).
The abscissa is the total flux from the annulus, i.e. the flux
per unit area times the total area of the annulus, which is chosen to
extend from $R/R_g=9.2$ to $10.4$.
}}

\vskip0.1in
\addtocounter{figure}{1}

In a subsequent study, Shimura \&~Takahara (1993) developed a more
accurate treatment of Comptonization.  They considered, 
as in the present paper, the Kompaneets form of the Compton scattering
emissivity, and solved the same list of structural equations
as in this paper (hydrostatic equilibrium, energy balance, and the
radiative transfer equation). However, they considered only free-free 
opacity, ignoring the many atomic opacity mechanisms,
and they assumed a two-stream
approximation for the radiative transfer (Rybicki \& Lightman 1979).
Unlike the present paper, they assumed separate energy balance equations
for the protons and the electrons.
However, they also showed that there was very little difference between  
the two temperatures, supporting our approach of having the same
kinetic temperature for all particles.


\subsection{Effects of Comptonization in hot disks}

We have constructed several disk models. We consider three
representative black-hole masses, $M = 10^6,\, 10^7,$ and $10^8$
M$_\odot$. For all three masses we take $L/{\rm L}_{\rm Edd} \approx 0.3$,
i.e.  mass accretion rates equal to 0.002, 0.02, and 0.2
M$_\odot$ yr$^{-1}$, respectively.  We assume a maximum rotation 
Kerr black hole.  For each disk, we consider two
values of the viscosity parameter $\alpha$: 0.1 and 0.01.
Analogously to Paper~III, we take
the individual annuli for the radial coordinates
$R/R_g$ (with $R_g=GM/c^2$ being the gravitational radius)
equal to 1.5, 2, 2.5, 3, 3.5, 4, 5, 6, 7, 8, 9, 20, 12, 15, 20,
25, 30, 40, 50 ,60, 70, 80, 90, 100, 150, 200, 250, 300, 400, 500,
600, 700, 800, 900, 1000, 1200, and 1500. A vertical structure model
is computed for all annuli for which the effective temperature
is higher than 5000 K; the cooler annuli are assumed to radiate
as black bodies.\footnote{We used the same prescription in Paper~III.  Note
that there were typos in that paper: the transition temperature was stated
to be 4000~K, whereas in fact it was chosen to be 5000~K.}~
The outer edge of the disk is chosen to be
the radius at which $T_{\rm eff}$ reaches 1000 K.

We first show the temperature structure for four representative annuli
of the disk with $M = 10^6$~M$_\odot$, $\dot M = 0.002$~M$_\odot$ yr$^{-1}$,
and $\alpha=0.1$ and $0.01$ -- see Figure~\FIGCOMA.

%

%
\hskip -0.2in
\parbox{3.0in}{\epsfxsize=3.5in \epsfbox{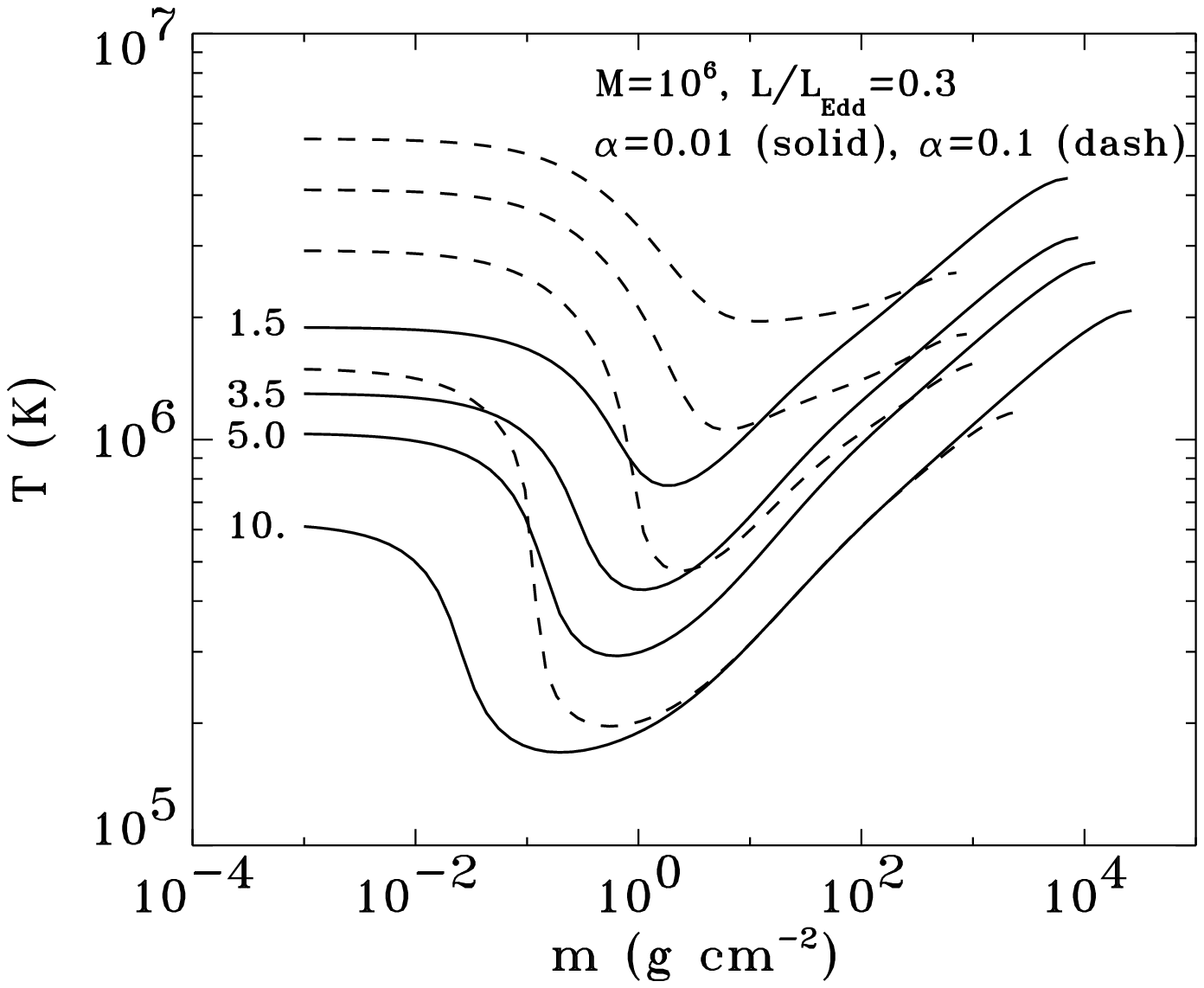}} \vskip 0.1in
\label{FIGCOMA}

\centerline{\parbox{3.5in}{\small {\sc Fig.~\FIGCOMA --} 
Temperature as a function of depth for
several representative annuli of the disk for $M = 10^6$~M$_\odot$,
$L/{\rm L}_{\rm Edd} \approx 0.3$; with $\alpha=0.01$ (solid lines), and
$\alpha=0.1$ (dashed lines).
From top to bottom, the curves correspond to radii $R/R_g =$
1.5, 3.5, 5, and 10.
}}

\vskip0.1in
\addtocounter{figure}{1}

The figure illustrates several interesting features. First, 
it shows the dependence of the surface temperature on both
effective temperature and total optical thickness, and thus on the
value of $\alpha$. We recall that the total thickness of the disk
is inversely proportional to $\alpha$ (Paper~III).
The hottest annulus has an effective temperature $\approx8 \times 10^5$~K.
The total optical thickness at the disk midplane is approximately
240 for the $\alpha=0.1$ disk, and 2400 for the $\alpha=0.01$ disk.
The mean photon destruction parameter for the $\alpha=0.1$ disk is 
about $10^{-6}$, while that for the $\alpha=0.01$ disk is about
$10^{-5}$. From Figure~\FIGSTB~we deduce that $T_0/T_{\rm eff}$ should be
around 6 and 2, respectively, for these two disks, which
is indeed verified by numerical calculations.

%

%
\hskip -0.2in
\parbox{3.0in}{\epsfxsize=3.5in \epsfbox{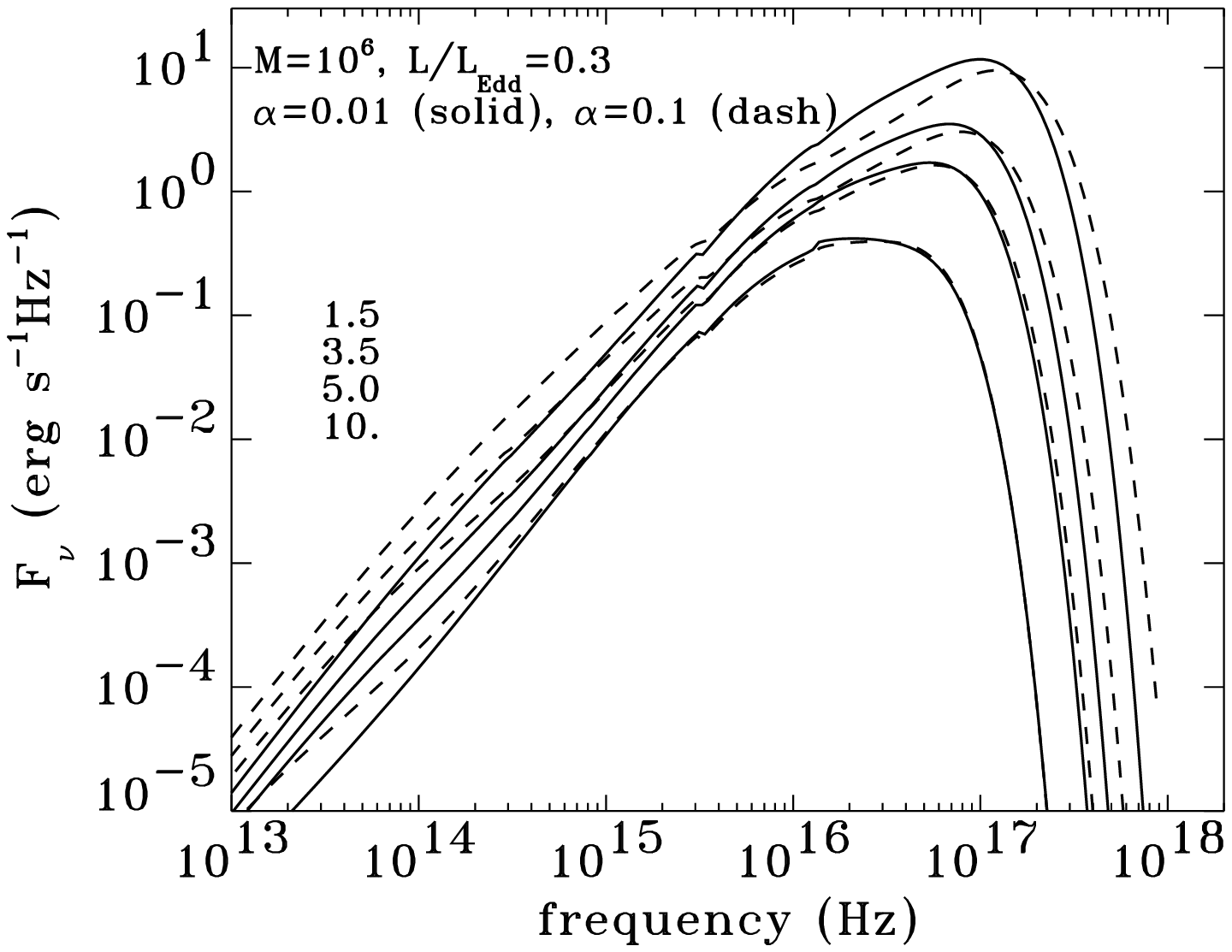}} \vskip 0.1in
\label{FIGCOMB}

\centerline{\parbox{3.5in}{\small {\sc Fig.~\FIGCOMB --} 
Local emergent flux for 
several representative annuli af the disk for $M = 10^6$~M$_\odot$,
$L/{\rm L}_{\rm Edd} \approx 0.3$; with $\alpha=0.01$ (solid lines), and
$\alpha=0.1$ (dashed lines).
The curves correspond to radii $R/R_g$ (from top to bottom)
1.5, 3.5, 5, and 10.
}}

\addtocounter{figure}{1}
\vskip0.1in

Finally, the corresponding emergent flux for the four annuli displayed
in Figure~\FIGCOMA~are shown in Figure~\FIGCOMB.  
Higher values of $\alpha$
have somewhat suppressed flux near the peak and enhanced flux at low
and high frequencies.  This effect is most pronounced at small radii,
i.e. lower total optical thickness.
To understand the behavior of the emergent flux, we display the effective
Compton $y$-parameter and other interesting quantities for the
annulus at $R/R_g=1.5$ of the disk with $M = 10^6$~M$_\odot$,
$L/{\rm L}_{\rm Edd} \approx 0.3$, and for the two values of $\alpha$
-- see Figure~\FIGCTHB.  This plot is analogous to Figure~\FIGC.

%

%
\hskip -0.2in
\parbox{3.0in}{\epsfxsize=3.5in \epsfbox{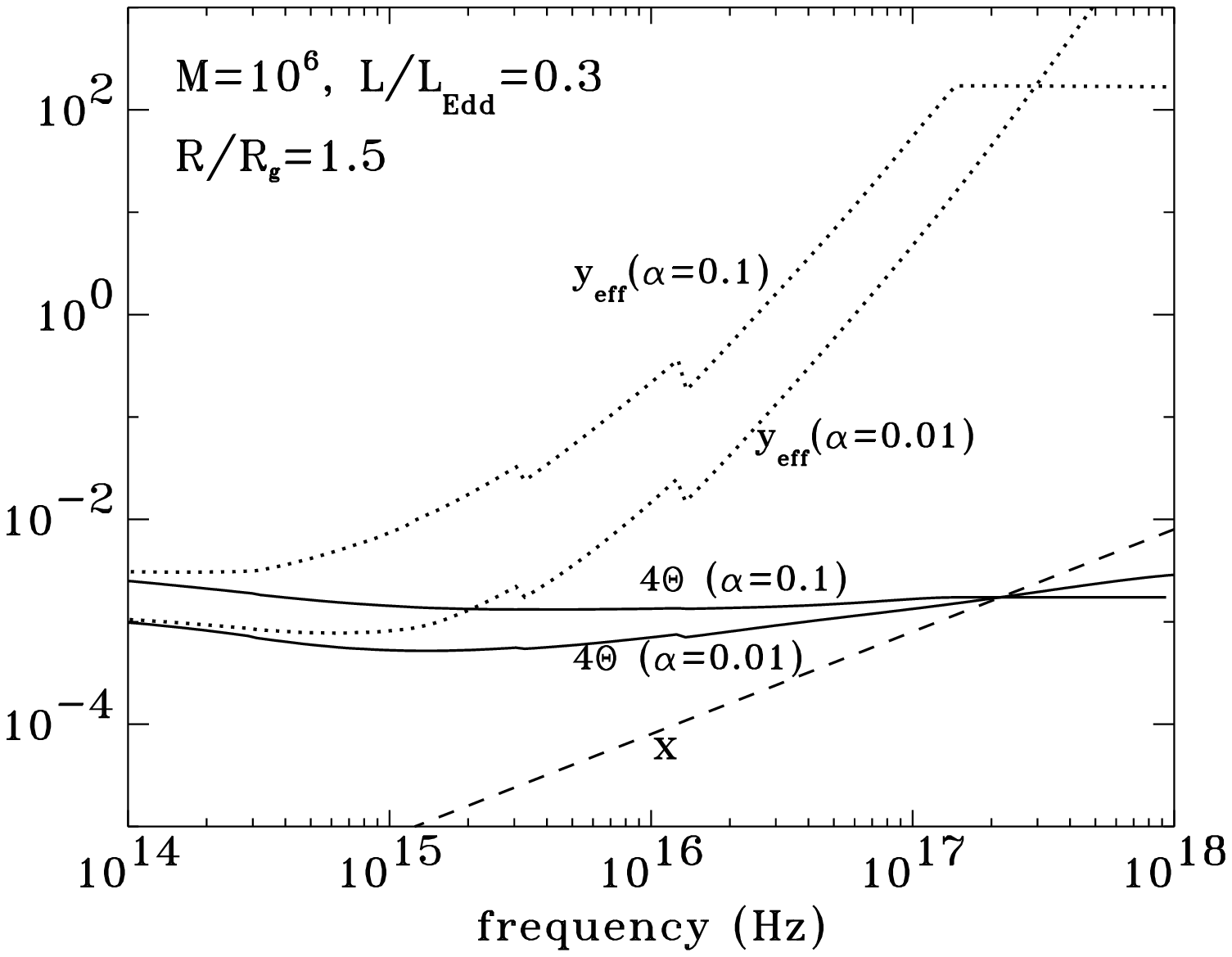}} \vskip 0.1in
\label{FIGCTHB}

\centerline{\parbox{3.5in}{\small {\sc Fig.~\FIGCTHB --} 
Computed values of $4 \Theta(T^\ast_\nu)$, $x$, and
the Compton parameter $y_{\rm eff}$ for the annulus at $R/R_g=1.5$
of the disk with $M = 10^6$~M$_\odot$, $L/{\rm L}_{\rm Edd} \approx 0.3$, 
and for $\alpha=0.01$ and 0.1. 
}}

\vskip0.1in
\addtocounter{figure}{1}

Increasing $\alpha$ decreases density in the disk, leading to greater
Comptonization because $y_{\rm eff} \propto n_{\rm e}^{-18/15}$ --
see equation (\ref{yeff2}). In line with this expectation,  $y_{\rm eff}$
in the models with $\alpha=0.1$ is more than an order of magnitude greater than
in the models with $\alpha=0.01$.  This is indeed nicely seen in 
Figure~\FIGCTHB.  The frequency where $4 \Theta(T^\ast_\nu) \approx x$
is about $2 \times 10^{17}$ Hz.  
However, the disk with $\alpha=0.1$ becomes effectively thin at 
$\nu \approx 1.3 \times 10^{17}$ Hz, while the annulus with $\alpha=0.01$
is effectively thick for all frequencies, so that 
$y_{\rm eff}$ keeps increasing with frequency.
In the model with $\alpha=0.1$, Comptonization is more efficient at
removing photons from the frequency range $\nu < 2 \times 10^{17}$~Hz to
the region around $2 \times 10^{17}$~Hz, so that the flux for
lower frequencies is depressed compared to the model with $\alpha=0.01$.
For higher frequencies, Comptonization is in contrast more efficient for
the model with $\alpha=0.01$, which leads to a larger shift of
photon energies towards lower frequencies,
thus resulting in a lower flux for the model with $\alpha=0.01$.


\subsection{Effects of metals}

We first computed a series of test models in order to
determine which metals must be treated in detail.
These models had the full array of metals as
discussed in Sect 2, i.e., H, He, C, N, O, Ne, Mg, Si, S, Ar,
Ca, Fe, and Ni. These test models showed that Mg, Si, S, Ca, and Ni
have only a marginal influence on models (both on the structure
and emergent spectral energy distribution), so we have computed
most models with only H, He, C, N, O, Ne, and Fe.

Figure~\FIGMETA~shows a comparison between models computed
assuming a simple H-He chemical composition and models with
metals. We plot here the total integrated spectrum
of the three representative disks.  We also show the integrated
spectrum assuming a black-body flux for each annulus.
As in our previous models, the self-consistent treatment of radiation
leads to a redistribution of flux from the optical and UV region
to the EUV and and soft X-ray region as compared to the integrated
black-body energy distribution.

The effect of metals is relatively small; the only
difference in the predicted spectra is seen in the high energy
tails.  Generally, the spectra begin to differ at a frequency
about three times higher than the frequency where the
emergent flux attains its maximum. 
The reason for this may be seen more clearly in Figure~\FIGCTHC,
where we present a comparison of the effective Compton $y$-parameter
for a representative H-He model and for a model with metals,
for a representative hot
annulus of the disk with $M = 10^6$~M$_\odot$, $L/{\rm L}_{\rm Edd} \approx
0.3$, and $\alpha=0.01$.  We do not show a comparison of local temperature 
for the models with metals to those of the H-He composition because
the differences are very small.  The metals thus do not influence the
emergent radiation through their effects upon the vertical structure,
but only through their effect on the thermal coupling parameter $\epsilon$
and consequently on the efficiency of Comptonization.

We see in Figure~\FIGCTHC~that in contrast to the H-He model,
the effective Compton $y$-parameter reaches a maximum in the region
$10^{17} \lta \nu \lta 2 \times 10^{17}$ because of the effects of metals,
in particular the O~VIII Lyman edge at $\nu \approx 2 \times 10^{17}$~Hz.
For higher frequencies, Comptonization is thus not very efficient.
The emergent flux is thus roughly given by the modified
black-body spectrum corresponding to the local temperature at the
effective depth of formation.  As we see in Figure~\FIGCTHC,
this temperature, which is proportional to $\Theta$, is decreased by
the influence of metals for $\nu \lta 2 \times 10^{17}$, and so
is the emergent flux.

%

%
\hskip -0.2in
\parbox{3.0in}{\epsfxsize=3.5in \epsfbox{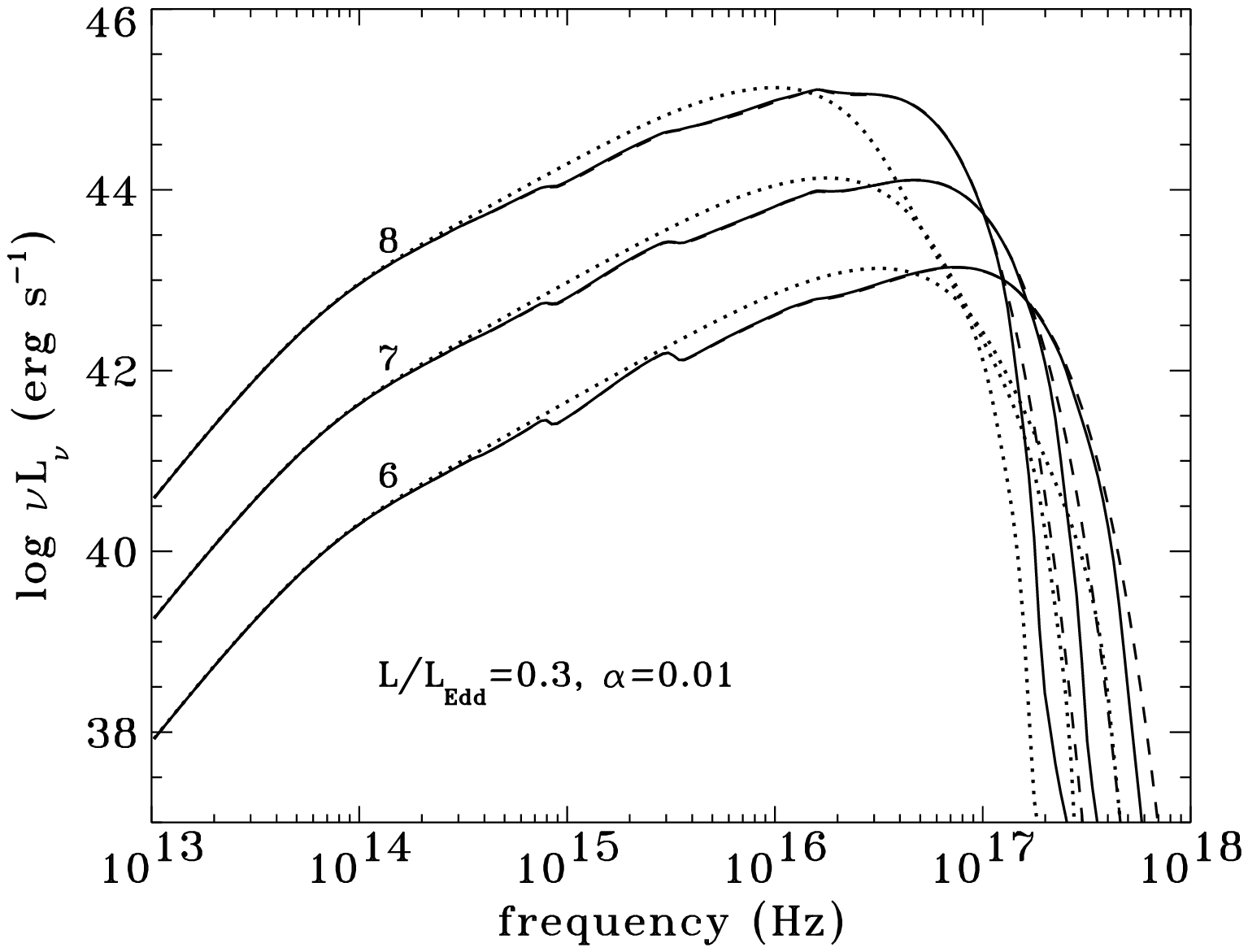}} \vskip 0.1in
\label{FIGMETA}

\centerline{\parbox{3.5in}{\small {\sc Fig.~\FIGMETA --} 
Integrated spectral energy distribution $\nu L_\nu$ for
three representative disks for $M = 10^6,\ 10^7,$ and $10^8$~M$_\odot$,
$L/{\rm L}_{\rm Edd} \approx 0.3$, $\alpha=0.01$. The solid lines show models
with metals, while the dashed lines show H-He models. Dotted lines
are the integrated energy distributions assuming the black-body
flux for each annulus.  The curves are labeled by the values of $\log M$.
}}

\vskip0.1in
\addtocounter{figure}{1}

%

%

%
\hskip -0.2in
\parbox{3.0in}{\epsfxsize=3.5in \epsfbox{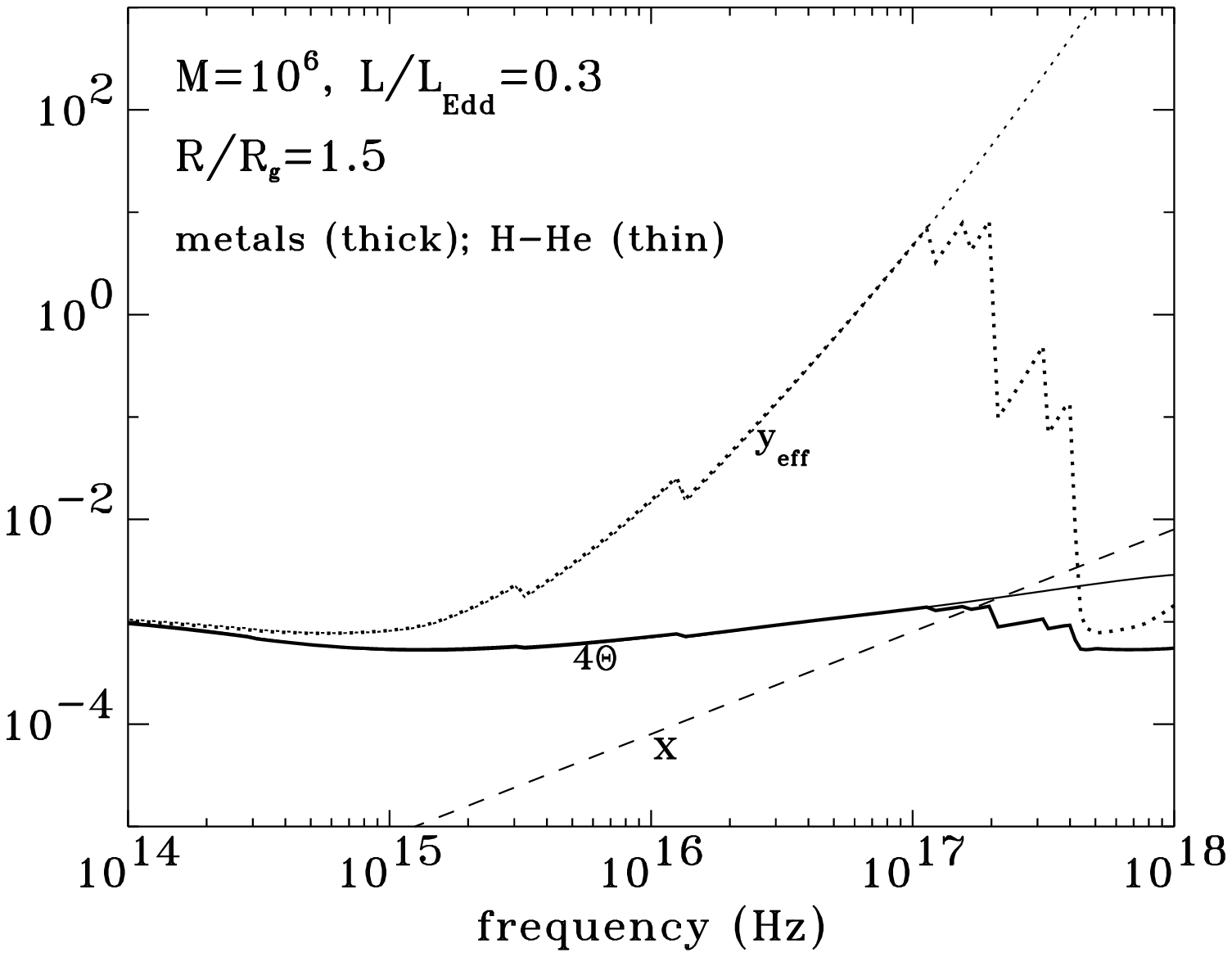}} \vskip 0.1in
\label{FIGCTHC}

\centerline{\parbox{3.5in}{\small {\sc Fig.~\FIGCTHC --} 
Computed values of $4 \Theta(T^\ast_\nu)$, $x$, and
the Compton parameter $y_{\rm eff}$ for the annulus at $R/R_g=1.5$
of the disk with $M = 10^6$~M$_\odot$, $L/{\rm L}_{\rm Edd} \approx 0.3$. 
and $\alpha=0.01$.  Thick lines represent the values for
the model with metals, while the thin lines represent the values for the
H-He model.
The influence of metal edges upon the
thermal destruction parameter $\epsilon$ and thus $y_{\rm eff}$ 
are clearly seen.  As is illustrated in Figure~\FIGMETB, the
discontinuities in the curve of $4 \Theta(T^\ast_\nu)$ correspond 
(with increasing frequency) to the Lyman edges of H~I, He~II, 
C~VI, N~VII, O~VIII, Ne~X, and Fe~XXIV.
}}

\addtocounter{figure}{1}
\vskip0.1in

In order to show the effects of metals more clearly, we plot
in Figure~\FIGMETC~the ratio of the integrated flux for models 
with metals to those for the H-He composition, 
for the same disks as shown in Figure~\FIGMETA.
The reason for a different shape of the curves for different
black hole masses is easy to understand.  In the present circumstances,
the metals contribute significantly to the opacity only through K-shell
ionization of C, N, O, and Ne, and L-shell ionization of Fe.
Only those species with edges well above $kT$ retain significant abundance.
Consequently, as the characteristic temperature rises, the mix of
heavy elements contributing to the high-frequency opacity shifts
to higher atomic number and therefore opacity at higher frequencies.
All edges appear blue-shifted in the integrated spectrum due to
relativistic effects.

%

%
\hskip -0.2in
\parbox{3.0in}{\epsfxsize=3.5in \epsfbox{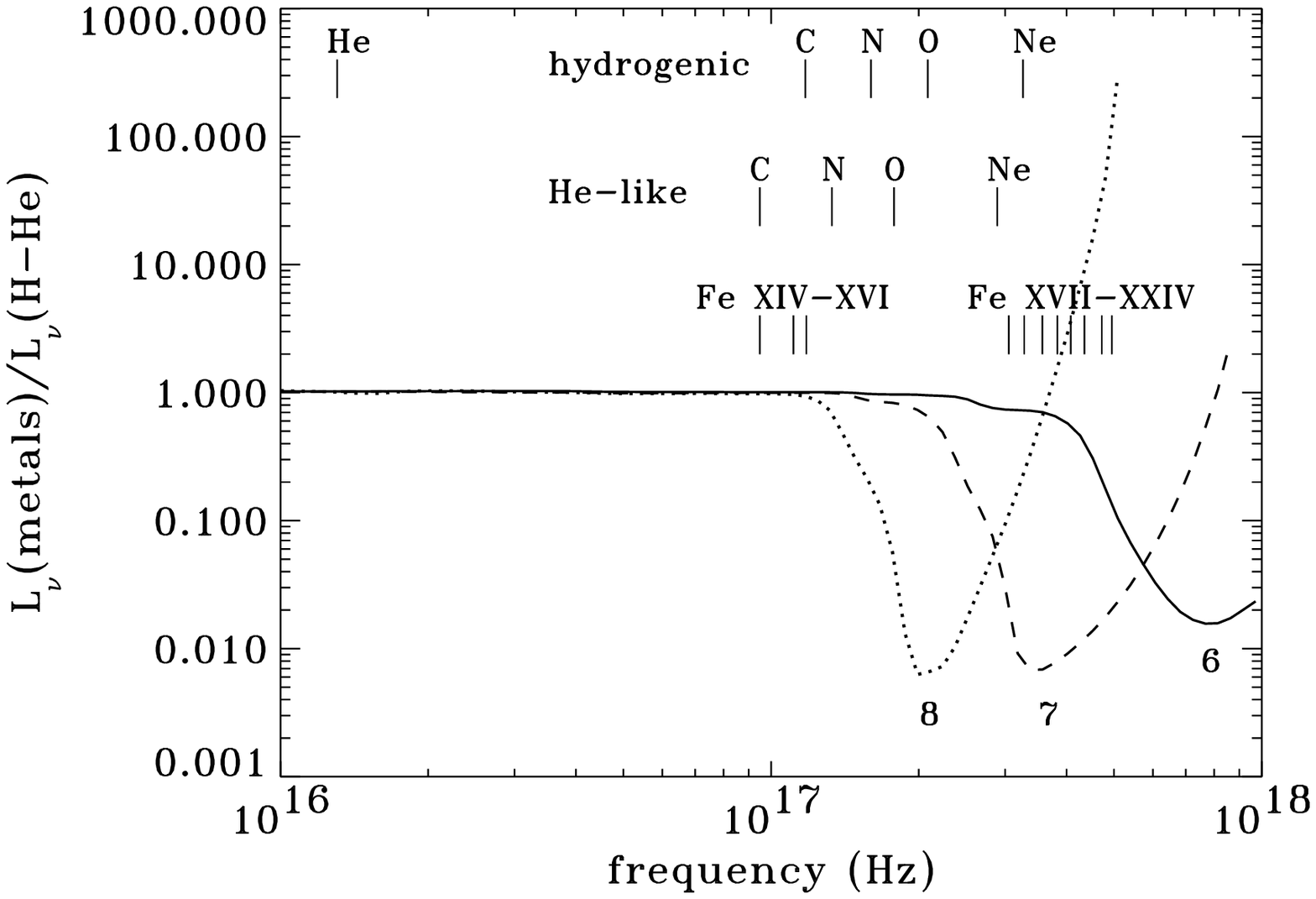}} \vskip 0.1in
\label{FIGMETC}

\centerline{\parbox{3.5in}{\small {\sc Fig.~\FIGMETC --} 
The ratio of integrated spectral energy distribution 
$L_\nu$ for the model with metals to that for the H-He model for
three representative disks with $M = 10^6,\ 10^7,$ and $10^8$~M$_\odot$,
$L/{\rm L}_{\rm Edd} \approx 0.3$, and $\alpha=0.01$.
The curves are labeled by the 
values of $\log M$.  We also show the positions of the photoionization
edges of the hydrogenic and He-like ions of He, C, N, O, and Ne, and
of various ions of iron.
}}

\vskip0.1in
\addtocounter{figure}{1}

The effects of changing the value of the viscosity parameter $\alpha$
are displayed in Figure~\FIGMETB. Here we display the same models
as in Figure~\FIGMETA, with metals, together with analogous models
for $\alpha=0.1$.  As before, we plot the black-body energy distribution 
as well.
The spectral energy distributions coincide for
$\nu < 10^{15}$~Hz; therefore only the high energy portions of the spectra
are shown.

%

%
\hskip -0.2in
\parbox{3.0in}{\epsfxsize=3.5in \epsfbox{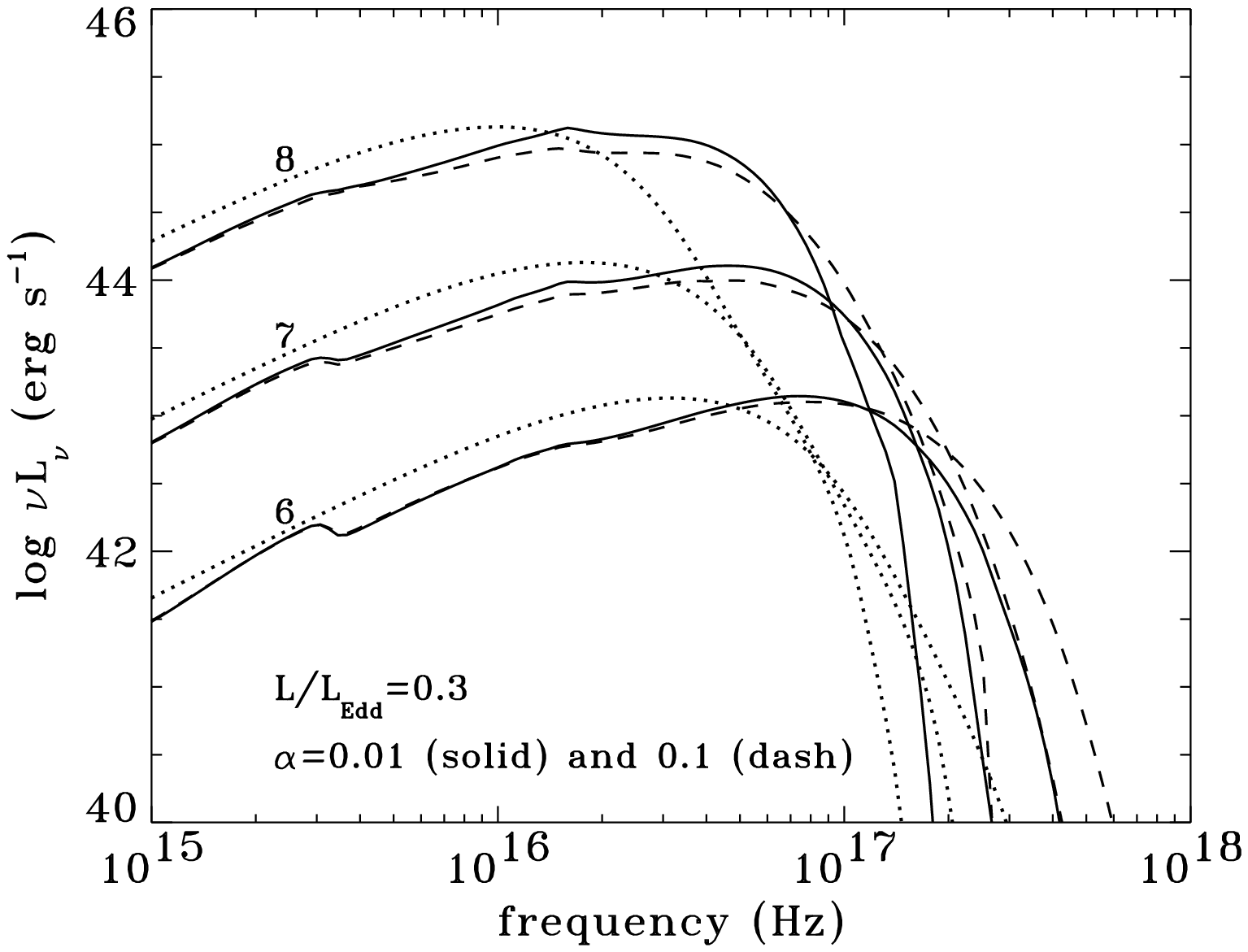}} \vskip 0.1in
\label{FIGMETB}

\centerline{\parbox{3.5in}{\small {\sc Fig.~\FIGMETB --} 
Integrated spectral energy distributions $\nu L_\nu$ for
three representative disks with $M = 10^6,\ 10^7,$ and $10^8$~M$_\odot$,
$L/{\rm L}_{\rm Edd} \approx 0.3$, computed with the influence of
metal continuum opacities. Solid lines show models with $\alpha=0.01$, while
dashed lines show models with $\alpha=0.1$. Dotted lines
are the integrated energy distributions assuming the black-body
flux for each annulus.
The curves are labeled by the values of $\log M$.
}}

\vskip0.1in
\addtocounter{figure}{1}

Models with $\alpha=0.1$ generally exhibit a lower flux at frequencies
below and around the maximum of flux, while the flux for high $\alpha$
models is much larger for the high-frequency region. 
This behavior is easily understood from the discussion of the individual
annuli presented in the Sect. 4.2.  Interestingly,
the flux for a model with $\alpha=0.1$ for a given $M$ is very 
close to the flux predicted for a model with $\alpha=0.01$ and $M$
about an order of magnitude lower!
We also conclude that for the hot disks, at least those
considered in the present paper, the effects of changing the 
value of the viscosity parameter $\alpha$ are {\em more important}
than the effects of metal continuum opacities.

Finally, as a reference point for later work that examines surface
ionization response to different heat deposition laws,
we show the ionization structure for three 
representative annuli of the disk with $M=10^6$~M$_\odot$, namely
the hottest annulus, $R/R_g=1.5$ with $T_{\rm eff} = 807,000$ K; 
an intermediate annulus, $R/R_g=15$ with $T_{\rm eff} = 207,000$ K;
and a ``cool'' annulus, $R/R_g=150$ with $T_{\rm eff} = 40,000$ K.
We show the ionization for two species: oxygen, which is 
a representative for the light metals; and for iron. 
The knowledge of ionization balance is important to show us what
are the most important opacity sources in the X-ray region.
The results are shown in Figure~\FIGIONA~for oxygen, 
and Figure~\FIGIONB~for iron.  The curves are labeled by the
effective charge of the ion, i.e., ``1'' denoting neutrals, 
``2'' once ionized, etc. The label appears close to the position 
of the maximum number density of the given ion.

Oxygen is essentially fully stripped in the hottest annulus 
throughout the whole vertical extent; while He-like oxygen
(O VII) dominates around $\tau \approx 1$ for the intermediate
annulus. In the cool annulus, O III and IV are dominant stages
around a Rosseland mean optical depth $\tau \approx 1$,
while O V and VI dominate deeper down. 
The ionization of iron is shown in Figure~\FIGIONB. The dominant
stages of iron are Fe XXII - XXV for the hot annulus; Fe XV - XVII
for the intermediate annulus, and Fe IV - VII for the cool annulus.


%
\hskip -0.2in
\parbox{3.0in}{\epsfxsize=3.5in \epsfbox{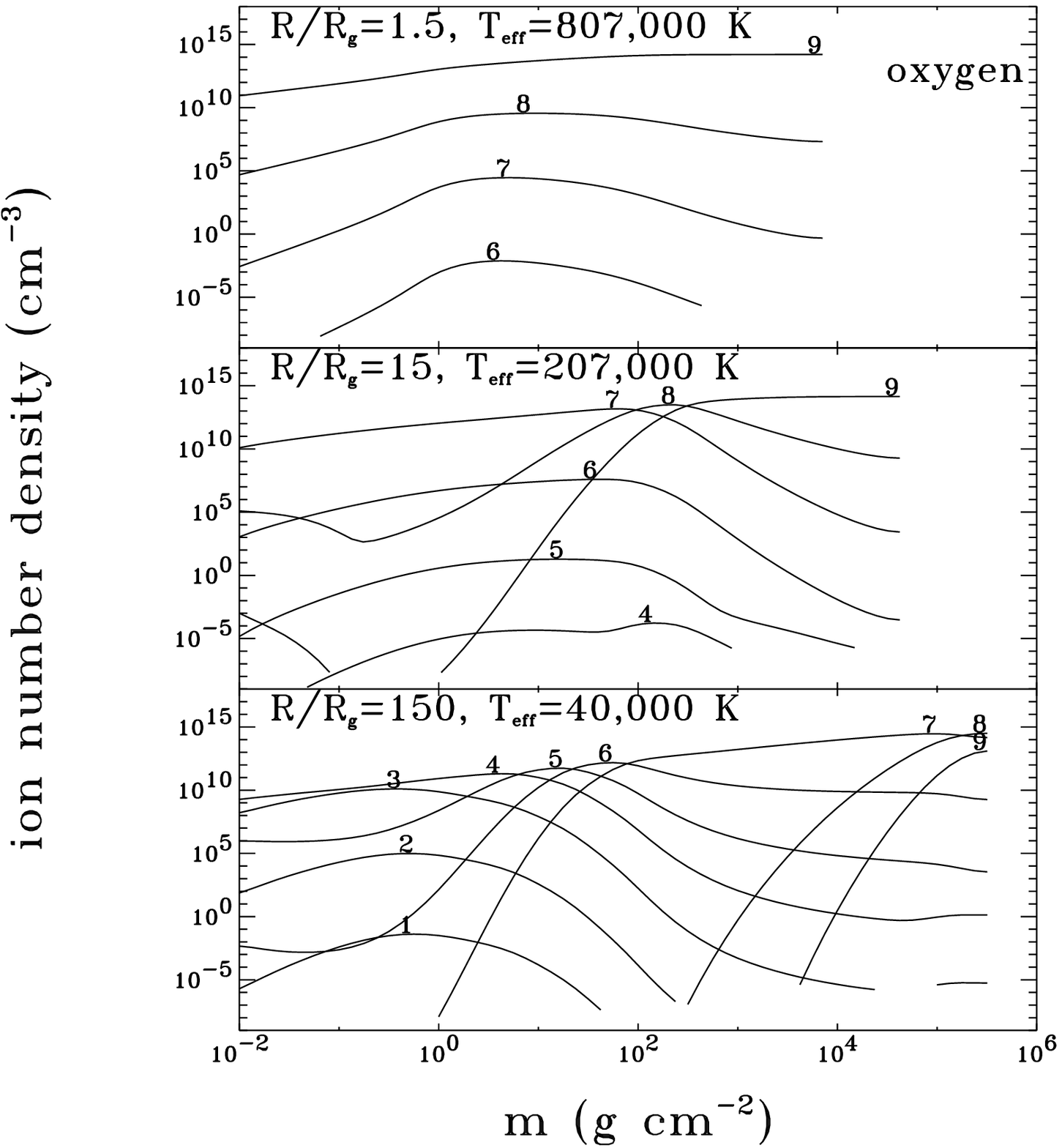}} \vskip 0.1in
\label{FIGIONA}

\centerline{\parbox{3.5in}{\small {\sc Fig.~\FIGIONA --} 
Ionization structure of oxygen for three
representative annuli of a disk model with $M = 10^6$~M$_\odot$,
$L/{\rm L}_{\rm Edd} \approx 0.3$, and $\alpha=0.01$, at radial distances
$R/R_g = 1.5,\, 15,$ and $150$. These annuli have effective
temperatures approximately 807,000; 207,000; and 40,000 K, respectively.
The abscissa is the column mass in g cm$^{-2}$ and the ordinate is the ion
number density in cm$^{-3}$.
The curves are labeled by the
effective charge of the ion, i.e., ``1'' denoting neutrals, 
``2'' once ionized, etc.
}}

\vskip0.1in
\addtocounter{figure}{1}

%


%
\hskip -0.2in
\parbox{3.0in}{\epsfxsize=3.5in \epsfbox{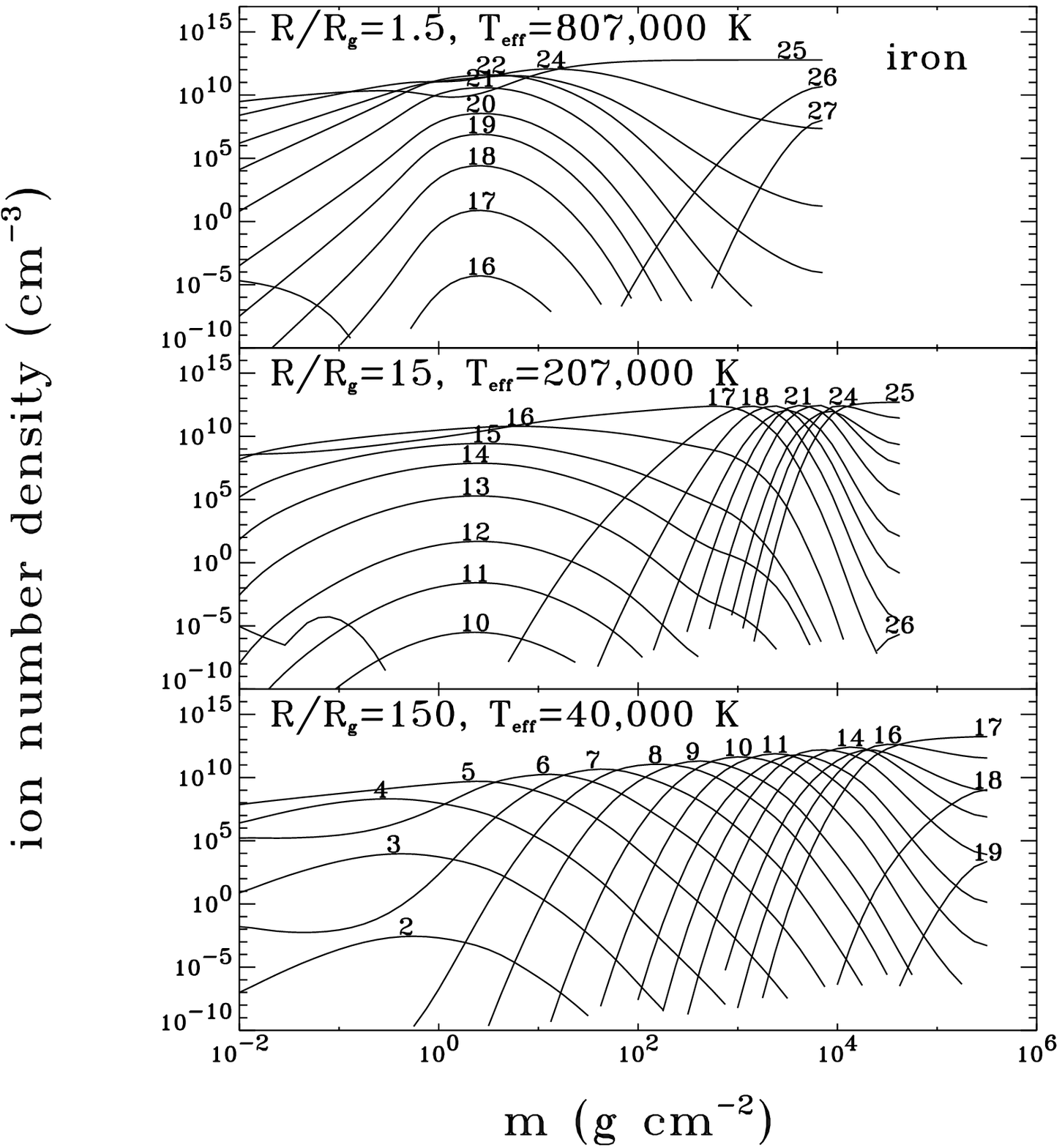}} \vskip 0.1in
\label{FIGIONB}

\centerline{\parbox{3.5in}{\small {\sc Fig.~\FIGIONB --} 
The same as in Figure~\FIGIONA, but for iron.
}}

\vskip0.1in
\addtocounter{figure}{1}

%


\section{Discussion}


\subsection{Effect of Comptonization on H and He Edges}

The observed absence of edge features at the hydrogen Lyman limit has been
a commonly cited problem with accretion disk models of AGN.  As we discussed
extensively in Paper~III, our disk models do not exhibit such features
for accretion rates which are reasonable fractions of the Eddington limit,
at least for the high black hole masses considered there.  Comptonization only 
reinforces this conclusion, because it tends to smear out flux
discontinuities (Czerny \& Zbyszewska 1991).  This is clear in the
spectra of individual annuli shown in Figure~\FIGFLB.  The effect is most
dramatic for the HeII Lyman edge, which occurs at the high frequencies where
Comptonization is most important, making it a much smoother feature (cf.
Figure~\FIGINT).

Our integrated disk spectra in  Figure~\FIGMETA~show that 
the H and He~II Lyman edges become more pronounced as the black 
hole mass is lowered.  This effect is
easily understood.  The vertically averaged density 
as a function of $R/R_g$ scales as (see Paper~II)
$\bar \rho  \propto (R/R_g)^{3/2} O(R/R_g) M\, \dot M^{-2} \alpha^{-1}$,
where the factor $O(R/R_g)$ accounts for relativistic corrections.
Since we consider models with fixed $L/L_{\rm Edd}$,  we have
$L/L_{\rm Edd} \propto \dot M/M = {\rm const}$, i.e., $\dot M \propto M$,
so that the density scales for our set of models as
$\bar \rho  \propto  M^{-1} \alpha^{-1}$.
The models with lower mass have higher density, and therefore higher
thermal coupling parameter $\epsilon$, so that the bound-free edges
are stronger.  The above considerations also indicate that the
edges are weaker for higher values of $\alpha$
(cf. Figure~\FIGMETB).
Our results
suggest that it might be easier to find broadened edge features in Seyferts
rather than quasars, although one would have to look out to reasonable
redshifts in order to avoid Galactic interstellar absorption.


\subsection{Metal Opacities and Soft X-ray Emission}

     In these disks, whose heating rate per unit mass is either constant or
declines toward the disk surface, the prime effect of heavy element K-shell
opacity is to partially suppress emission in the soft X-ray bands affected
by these photoionization edges.  The particular edges that most
powerfully affect the spectrum depend on the central mass, rising in
energy as the central mass decreases (Figure~\FIGMETC).  To a very rough
approximation, the central energy of the edge that produces the greatest
X-ray absorption scales as $M^{-1/4}$, similarly to the temperature
scale in the disk.  As a result of this additional opacity,
flux is redirected to frequency
bands with smaller opacity in the EUV (Figure~\FIGMETA).  

     However, there is reason to think this effect may be strongly
model-dependent.  As shown in Figure~\FIGMETB, its magnitude can be
substantially altered by changes in the local column density (as
parameterized, for example, by the stress parameter $\alpha$).  Thinner
disks present less soft X-ray opacity, and therefore show weaker breaks
in that region than do thicker disks.
Similarly, if the heating rate per unit mass were to increase, rather
than decrease, in the surface layers, these same features might swing into
emission.

Observations of quasar spectra in the $\sim100-1000$~eV band reveal a
complicated (and possibly contradictory) story.  Low redshift radio-quiet
quasars appear to be consistent with single steep power-laws with an average
spectral slope of $d\ln L_\nu/d\ln\nu\simeq-1.7$ (Laor et al. 1997).
On the other hand, in many cases the spectra of lower luminosity nearby
radio quiet AGN, while following power laws above $\simeq 2$~keV, rise
sharply at lower energies (Turner \& Pounds 1989), producing a so-called
``soft excess''.  Similarly, Blair et al. (2000) find that the composite
X-ray spectra of quasars binned over various redshift intervals are best
fit by a combination of a power-law and single-temperature blackbody.  The
power law component has a spectral slope of $d\ln L_\nu/d\ln\nu\simeq-0.9$,
consistent with the 2-10~keV slope seen in nearby Seyferts.  The blackbody
component that they fit has a temperature that rises from $\simeq100$~eV
at $z\simeq0$ to $\simeq270$~eV at $z\simeq2.5$.  The luminosity of
this component is $\simeq10^{44}$~ergs~s$^{-1}$, independent of redshift.

As shown in Figure~\FIGMETB, the spectral energy distributions expected 
from our disk models in the 0.2-2~keV ($5\times10^{16} -5\times10^{17}$~Hz)
band show a broad range of slopes.  At the highest energies they all
show exponential rollovers, however, so the hard power-law spectra observed
in AGN require a separate component.  This is widely believed to be due
to thermal Comptonization in a hot corona.  A modification of our models
to include increased viscous dissipation near the surface might be able
to self-consistently produce such a corona, and we will explore this
in a future paper.

Whether or not the models presented in this paper can provide good fits
to the soft excess alone is unclear.  Our models require $L/{\rm L_{Edd}}<0.3$
in order to be consistent with our assumption that the disk is geometrically
thin.  A black hole mass of at least $10^8$~M$_\odot$ is required to produce
observed quasar luminosities, and Figure~\FIGMETB~shows that our accretion disk
models for such black holes will produce spectra that peak below 200~eV.
This might nevertheless be consistent with the 270~eV temperatures of the
Blair et al. blackbody components at high redshift.  Our models are not
well-fit by single temperature blackbodies near the peak, so detailed
spectral fits are required.

Huge soft X-ray excesses are a hallmark of narrow line Seyfert 1 galaxies
(Boller, Brandt, \& Fink 1996).  In order to explain the soft X-rays,
accretion disk models of these sources are inevitably forced to high
accretion luminosities (Eddington or above), if they are based on the
assumption of local blackbody emission (e.g. Mineshige et al. 2000).
However, our models shown in Figure~\FIGMETB~indicate that these high accretion
rates may not in fact be necessary.  Comptonization extends the emission
well beyond that expected from a multi-temperature blackbody model of the
same disk.  This might contribute substantially to the observed soft X-ray
excesses in narrow line Seyfert 1s, particularly at the relatively
low black hole masses that might exist in these sources.


\subsection{Radiative vs. Convective Heat Transport}

We have assumed throughout the present paper that heat is transported vertically
through the disk by radiative transfer.  However, it turns out that all our
disk models are convectively unstable.  Away from ionization zones, the square
of the Brunt-V\"ais\"al\"a frequency in the optically thick limit may be
written as
\begin{equation}
N^2=g{d\ln\rho\over dz}\left({3P\over 5P_{\rm gas}+4P_{\rm rad}}\right)
\left[{d\ln P\over d\ln\rho}-\left({d\ln P\over d\ln\rho}\right)_{\rm ad}
\right],
\label{eqbrunt}
\end{equation}
where the adiabatic gradient is
\begin{equation}
\left({d\ln P\over d\ln\rho}\right)_{\rm ad}={5P_{\rm gas}+4P_{\rm rad}
\over3P}\, ,
\end{equation}
and $P=P_{\rm rad}+P_{\rm gas}$ is the total (radiation plus gas) pressure.
Note that the adiabatic gradient is $4/3$ for a radiation pressure dominated
medium, but $5/3$ for a gas pressure dominated medium.
Equation (\ref{eqbrunt}) shows that convective instability occurs when
$d\ln P/d\ln\rho$ exceeds the adiabatic
gradient, with a characteristic growth rate of order the orbital frequency.
Figure~\FIGCONV~
depicts ${d\ln P/d\ln\rho}$ as a function of column density
for a number of annuli in our $M=10^6$~M$_\odot$, $L=0.3$~L$_{\rm Edd}$,
$\alpha=0.01$ model.  We see that all our annuli are convectively unstable
at high column densities.  Photon diffusion is not included in equation
(\ref{eqbrunt}), but the effect of a long photon mean free path near the
photosphere would be to diminish the radiation pressure contribution to
the adiabatic gradient, which therefore approaches $5/3$.


%
\hskip -0.2in
\parbox{3.0in}{\epsfxsize=3.5in \epsfbox{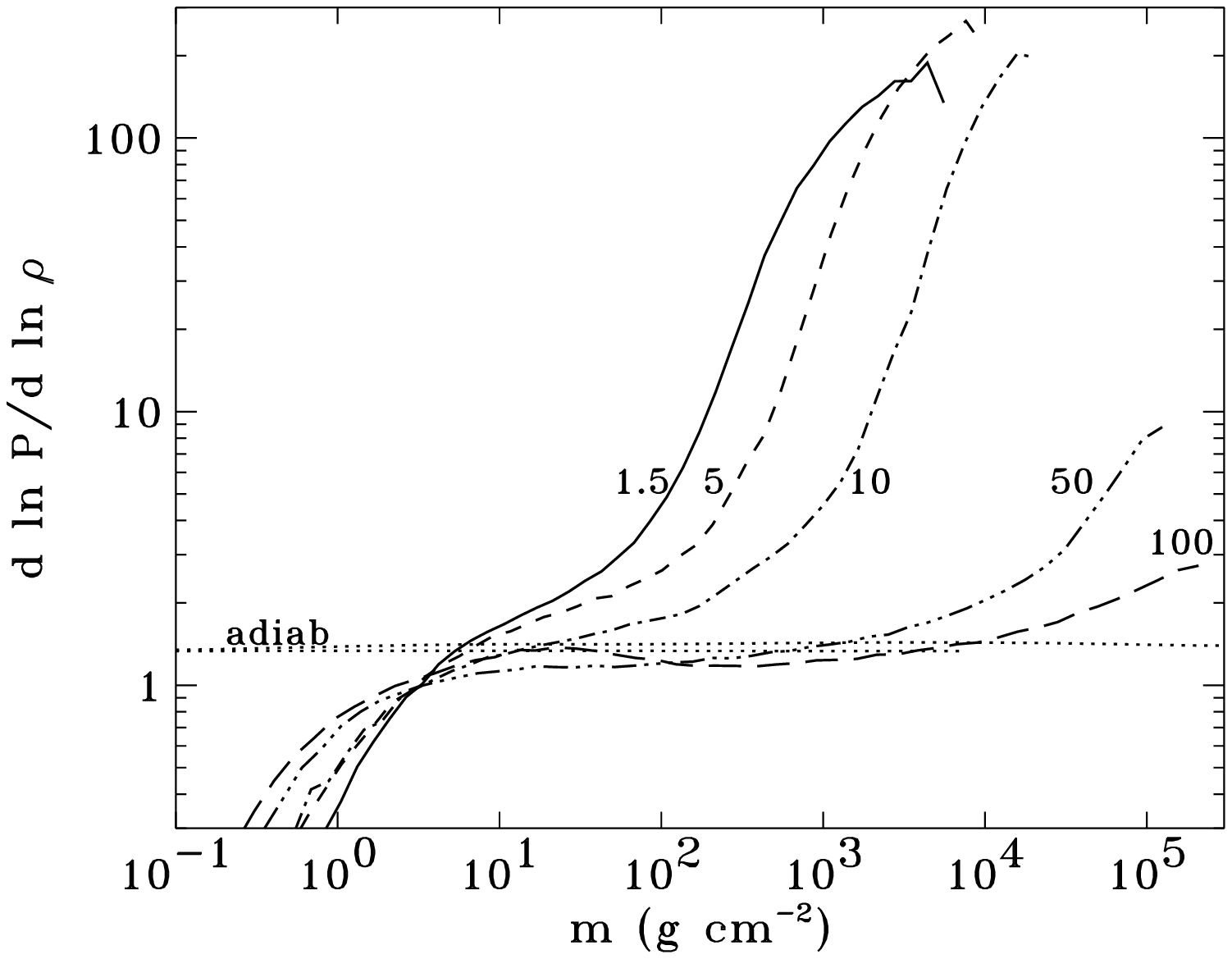}} \vskip 0.1in
\label{FIGCONV}

\centerline{\parbox{3.5in}{\small {\sc Fig.~\FIGCONV --} 
Plot of the logarithmic gradient ${d\ln P/d\ln\rho}$ 
as a function of column density for a number of annuli in the 
$M=10^6$~M$_\odot$, $L=0.3$~L$_{\rm Edd}$, $\alpha=0.01$ model. 
The curves are labeled by the value of $R/R_g$.  Dotted lines
show the adiabatic gradient for the hottest annulus ($R/R_g=1.5$  -- lower
curve), and for the coolest annulus ($R/R_g= 100$ -- upper curve).
In both cases, the adiabatic gradient is close to $4/3$, although
it is higher for the cooler model, where the gas pressure makes
a non-negligible contribution.
}}

\vskip0.1in
\addtocounter{figure}{1}

These instabilities depend of course on our assumed vertical dissipation
profile.  At the present time this profile is unknown, although numerical
simulations of the nonlinear development of the magneto-rotational instability
are beginning to address this question, at least in gas pressure dominated disks
(e.g. Miller \& Stone 2000).  It is nevertheless clear that our models are not
self-consistent, and convective heat transport should be included.  
How this will affect the spectrum is unclear.  While the
scattering photospheres appear to be convectively stable, the continuum is
formed in deeper, potentially unstable layers.  More seriously, hydrodynamical
simulations show that convective heat transport changes the overall heat
content in the disk, altering the scale height and surface gravity of the
photosphere (Agol et al. 2000).  These simulations also produce an
approximately isentropic configuration (at least when horizontally averaged),
similar to what is seen in convection zones in stars.
Assuming that magnetic fields do not change things (see, however,
Hawley, Balbus, \& Stone 2001), the assumption of constant entropy may
provide a good first approximation to convective heat transport in unstable
zones.  We will address convective heat transport in disks in a future paper.



\section{Conclusions}

We have extended our modeling procedure described in the previous papers
of this series in two respects.
First, we have self-consistently included the effects of
Compton scattering, both on the disk structure and on the emergent
radiation.  Second, we have included the effects of continuum opacities
of the most important metals (C, N, O, Ne, Mg, Si, S, Ar, Ca, Fe, Ni).

The present models are thus based on the following assumptions:\\ 
(i) The energy is generated
by turbulent viscous dissipation, with the vertically-averaged viscosity 
described through the Shakura-Sunyaev parameter $\alpha$.\\
(ii) The kinematic viscosity is constant with depth.\\
(iii) Convection and conduction are neglected.\\ 
(iv) No external irradiation or self-irradiation of the disk is 
considered.\\ 
(v) Effects of line opacity are neglected.\\ 
The underlying assumption, not listed here, is that the 1-D approach
is appropriate, i.e., that the disk may be described as a set of
mutually non-interacting, concentric annuli.

We have also developed a simple analytic model that provides
an estimate of the vertical temperature structure of the disk in the
presence of Comptonization. An interesting by-product of this
model is an estimate of the surface temperature. In particular, 
we have shown under which conditions a hot outer layer 
(a ``corona'') is formed.

We have recomputed a number of models from the very large
central-mass grid of 
Hubeny et al. 2000 (Paper~III) with Comptonization taken into
account, and showed 
that the predicted disk-integrated spectra differ only modestly
from the previous models.   We have concluded that
the grid of models from Paper~III is not significantly plagued by
neglecting Comptonization, except perhaps for the very hottest
(smallest central mass) models
of the sample. We have also compared our results to analogous models
of Ross, Fabian, \&~ Mineshige (1992).
The agreement of the computed continuum flux
is very good, although there are some differences,
particularly at the shortest and the longest wavelengths. 
The predicted profiles for the hydrogen and helium lines differ
significantly, which is very likely explained by differences
in treating the lines, in particular by the escape probability
treatment adopted by Ross et al. 
 
We have also computed a set of vertical structure models for
black hole masses of $10^6$, $10^7$, and $10^8$~M$_\odot$,
total luminosity $L/L_{\rm Edd} \approx 0.3$, and for two
values of the $\alpha$ parameter: 0.1, and 0.01. Unlike the previous
papers of this series, we consider a depth-independent kinematic
viscosity. The main effect of Comptonization in this mass
range is to transfer about 10 percent of the total luminosity
from the high energy tail (above $6\times10^{16}$ Hz) to the
peak of the spectrum (between 1.5 and $4\times10^{16}$ Hz.)
We have computed a number of tests, and found that for
the physical conditions studied in this paper (temperatures up
to $10^7$ K), the effects of Mg, Si, Ca, and Ni are negligible.
We have shown that the effects of other metals are still relatively modest.
The only significant differences in the predicted spectra
occur in the high-frequency tails
of the spectra, where the flux is already rather low.
Interestingly, the effects of changing the viscosity parameter $\alpha$ 
from 0.01 to 0.1 are found to be larger that the effects of metals.
Models with $\alpha=0.1$ generally exhibit a lower flux at frequencies
below and around the maximum of flux, while the flux for high $\alpha$
models is much larger for the high-frequency region.

The effects studied in this paper are most dramatic in the EUV/soft X-rays.
As shown in Figure~\FIGMETB, this high frequency region of the spectrum
differs enormously from that expected from
multitemperature blackbodies.  Fits to the soft X-ray excesses of AGN which
assume local blackbodies will necessarily require an artificially high
accretion rate and accretion luminosity.

In future papers of this series, we will relax assumption (ii),
and will compute models that allow for an increase of the heating rate
(kinematic viscosity) with height, leading to the formation of very hot upper
layers of the disk.  
We will also relax assumptions (iii), incorporating convective heat
transport, and (v), studying the effects of 
line opacities on the disk structure in detail.


\acknowledgments

We thank Randy Ross for providing us with his model in digital
form. We also thank Tim Kallman for sharing with us his collection
of atomic data.
This work was supported in part by NASA grant NAG5-7075.


}
\appendix

\section{Appendix A: The Compton Scattering Source Function}

We treat Comptonization in the non-relativistic, diffusion limit, where
the exact integral over the redistribution function is approximated
by a Kompaneets type operator.  The general form of the radiative transfer
equation for arbitrary geometries in this limit has been derived by
Babuel-Peyrissac \& Rouvillois (1969) and Nagirner \& Poutanen
(1993).\footnote{The equations in these papers contain small typos which we
have corrected.}  In the plane-parallel geometry considered here, the
radiation field is axisymmetric, but still has a dependence on the polar
direction cosine $\mu$.  After integrating over azimuthal angles, we find
that the contribution of Compton scattering to the right-hand side of the
radiative transfer equation (i.e.  sources minus sinks) is
\begin{equation}
\label{com_gen}
\mu\, \left( {\partial I \over \partial  z} \right)^{\rm Compt} \!\! =
{3 \over 16}\, n_e \sigma_{\rm T} \int_{-1}^1 d\mu_1\, \left[ (I_1 - I)(A - 2B + C)
+ B(D+E) \right] \, .
\end{equation}
For simplicity of notation here, we have dropped the frequency subscript,
$\nu$. The specific intensity
$I\equiv I_\nu(\mu)$, while $I_1 \equiv I_\nu(\mu_1)$.

Here,
\begin{equation}
A(\mu,\mu_1) = \alpha \, ,
\end{equation}
\begin{equation}
B(\mu,\mu_1) = x \, (\alpha - \beta)\, ,
\end{equation}
\begin{equation}
C(\mu,\mu_1) = 2 \Theta \, (8 - 8 \gamma - 3 \alpha + 2 \beta)\, ,
\end{equation}
\begin{equation}
D = {h\nu^4 \over k T} \left[ {k^2 T^2 \over h^2}
{\partial^2 \over \partial \nu^2}\left( {I_1 \over \nu^3} \right) +
{k T \over h}{\partial \over \partial \nu}\left( {I_1 \over \nu^3} \right)
\left(1 + {c^2 \over h \nu^3}\, I \right) \right]\, ,
\end{equation}
and
\begin{equation}
E = 4 \left[ 
{k T \over h}\nu^3 {\partial \over \partial \nu}\left( {I_1 \over \nu^3} \right)
+ I_1 \left(1 + {c^2 \over 2 h \nu^3}\, I  \right) \right]\, ;
\end{equation}
where
\begin{equation}
x = {h\nu \over m_e c^2}\, , \quad\quad  \Theta = {kT \over m_e c^2}\, ,
\end{equation}
and where the matrices $\alpha$, $\beta$, and $\gamma$ are
given by
\begin{equation}
\alpha(\mu,\mu_1) = 3 \mu^2\mu_1^2 - \mu^2 - \mu_1^2 + 3,
\end{equation}
\begin{equation}
\beta(\mu,\mu_1) = 5\mu \mu_1 +5 \mu^3 \mu_1^3 - 3 \mu^3 \mu_1 - 3 \mu \mu_1^3,
\end{equation}
\begin{equation}
\gamma(\mu, \mu_1)= \mu \mu_1\, ,
\end{equation}
so that $\alpha$ is an even function of $\mu$ or $\mu_1$, while $\beta$
and $\gamma$ are odd.

The radiative transfer equation is conveniently written by splitting
the total absorption/emission term into four parts:
absorption ($\sigma_\nu I_\nu$), spontaneous emission ($\eta_\nu$), 
stimulated emission ($\eta_\nu^{\rm stim} I_\nu$), and thermal 
absorption/emission ($\kappa_\nu^{\rm th}$, $\eta_\nu^{\rm th})$,  i.e.,
\begin{equation}
\mu\, {\partial I_\nu \over \partial  z} = -(\kappa_\nu^{\rm th}+\sigma_\nu) 
I_\nu 
+ \eta_\nu +\eta_\nu^{\rm th} + \eta_\nu^{\rm stim} I_\nu,
\end{equation}
where
\begin{equation}
\sigma_\nu = {3 \over 16}\, n_e \sigma_{\rm T} \int_{-1}^1 d\mu_1\, (A - 2B + C),
\end{equation}
\begin{equation}
\eta_\nu = {3 \over 16}\, n_e \sigma_{\rm T} \int_{-1}^1 d\mu_1\, 
[ I_1 \, (A - 2B + C) + B(D^\prime + E^\prime) ],
\end{equation}
\begin{equation}
\eta_\nu^{\rm stim} = {3 \over 16}\, n_e \sigma_{\rm T} \, {c^2 \over 2h\nu^3}
\int_{-1}^1 d\mu_1\, B\,  \left[ {2 h\nu^4 \over kT} {kT \over h} 
{\partial \over \partial\nu}\left( {I_1 \over \nu^3} \right) + 4 I_1 \right]\, ,
\end{equation}
where $D^\prime$ and $E^\prime$ are the $D$ and $E$ terms without the
stimulated emission contributions, i.e.,
\begin{equation}
D^\prime = {h\nu^4 \over k T} \left[ {k^2 T^2 \over h^2}
{\partial^2 \over \partial \nu^2}\left( {I_1 \over \nu^3} \right) +
{k T \over h}{\partial \over \partial \nu}\left( {I_1 \over \nu^3} \right)
\right]\, ,
\end{equation}
and
\begin{equation}
E^\prime = 4 \left[ 
{k T \over h}\nu^3 {\partial \over \partial \nu}\left( {I_1 \over \nu^3} \right)
+ I_1 \right]\, .
\end{equation}
We introduce the following notation
\begin{equation}
I_\nu^\prime = {\partial I_\nu\over \partial \ln \nu}\, , \quad\quad
I_\nu^{\prime\prime} = {\partial^2 I_\nu\over \partial (\ln \nu)^2}\, .
\end{equation}
so the $D^\prime$ and $E^\prime$ terms are written
\begin{equation}
D^\prime = {kT \over h\nu} (I_1^{\prime\prime} - 7 I_1^\prime + 12 I_1)
+ (I_1^\prime - 3 I_1)\, ,
\end{equation}
\begin{equation}
E^\prime = 4\,  {kT \over h\nu}\,  (I_1^\prime - 3 I_1) + 4\,  I_1 \, , 
\end{equation}
and thus
\begin{equation}
x (D^\prime +E^\prime) =  \Theta  \, I_1^{\prime\prime} + 
(x-3\Theta )\, I_1^\prime + x \, I_1 \, . 
\end{equation}
After some algebra, we obtain for the individual terms
\begin{equation}
\sigma_\nu = n_e \sigma_{\rm T} \, (1 - 2 x)\, ,
\end{equation}
\begin{eqnarray}
\eta_\nu = {3 \over 16} n_e \sigma_{\rm T} \int_{-1}^1 d\mu_1 \{
\left[ (1-x-6\Theta ) \alpha  + 16\Theta  + (x+4\Theta ) \beta  - 16\Theta  
\gamma  \right] I_1
\nonumber \\
+ (\alpha - \beta) (x-3\Theta ) I_1^\prime + (\alpha - \beta) \Theta  
I_1^{\prime\prime} \}
\, ,
\end{eqnarray}
and
\begin{equation}
\eta_\nu^{\rm stim} = {3 \over 16} n_e \sigma_{\rm T} \, {c^2 \over 2h\nu^3}
\int_{-1}^1 d\mu_1 2 x \, (\alpha - \beta) \, (I_1^\prime - I_1)\, .
\end{equation}
The thermal terms $\kappa_\nu^{\rm th}$ and $\eta_\nu^{\rm th}$
contain bound-free, free-free, and possibly also 
bound-bound contributions from all transitions which are taken into account.
The stimulated emission is treated as negative absorption, as is customary
in astrophysical radiative transfer.

The Compton emission terms are more complicated than the thermal emission
term $\eta_{\nu}^{\rm th}$ because they depend explicitly on angle.
To write them in a more useful form, let us first introduce the 
traditional moments of the specific intensity, defined by
\begin{equation}
[J, \, H, \, K,\, N] \equiv {1 \over 2} \int_{-1}^1 d\mu_1 \, I_1 \cdot 
[1,\, \mu_1,\, \mu_1^2,\, \mu_1^3] .
\end{equation}
and also special ``pseudo-moments'' of the specific intensity with 
weights $\alpha$, $\beta$, and $\gamma$, defined as
\begin{equation}
[{\cal J}_\alpha, \, {\cal J}_\beta, \, {\cal J}_\gamma ] \equiv
{3 \over 16}  \int_{-1}^1 d\mu_1 \, I_1 \cdot 
[\alpha, \, \beta, \, \gamma],
\end{equation}
The pseudo-moments can be expressed through the ordinary moments as
\begin{equation}
\label{mom_alp}
{\cal J}_\alpha = {3 \over 8} \left[ (3 -\mu^2)\, J + (3\mu^2 - 1)\, K
\right],
\end{equation}
\begin{equation}
{\cal J}_\beta = {3 \over 8} \, \mu \left[ (5 - 3\mu^2)\, H + (5\mu^2 - 3)\, N
\right],
\end{equation}
\begin{equation}
{\cal J}_\gamma = {3 \over 8} \, \mu \, H .
\end{equation}
Because the pseudo-moments are integrals over matrices $\alpha$, $\beta$, and
$\gamma$, they are explicit functions of the angle, $\mu$. We may then
also introduce moments of ${\cal J}_\alpha$, ${\cal J}_\beta$, and
${\cal J}_\gamma$. The zero-order moments are given by
\begin{equation}
{1\over 2} \int_{-1}^1 {\cal J}_\alpha \, d\mu = J \, ,
\end{equation}
and (since ${\cal J}_\beta$ and ${\cal J}_\gamma$ are antisymmetric functions
of $\mu$),
\begin{equation}
{1\over 2} \int_{-1}^1 {\cal J}_\beta \, d\mu =  
{1\over 2} \int_{-1}^1 {\cal J}_\gamma \, d\mu = 0\, .
\end{equation}
The first-order moments are given by
\begin{equation}
{1\over 2} \int_{-1}^1 {\cal J}_\alpha \, \mu \, d\mu = 0 \, ,
\end{equation}
and 
\begin{equation}
\label{mom_bg}
{1\over 2} \int_{-1}^1 {\cal J}_\beta\,  \mu \, d\mu =  {2\over 5}\, H\, ,
\quad\quad
%
{1\over 2} \int_{-1}^1 {\cal J}_\gamma\,  \mu \, d\mu = {1\over 8}\, H .
\end{equation}
%

Using the definitions of the optical depth, $\tau_\nu$, the photon
destruction probability, $\epsilon_\nu$, and the scattering probability,
$\lambda_\nu$, introduced in section 2.1, we obtain for the full,
angle-dependent transfer equation
\begin{equation}
\label{rtec_ang}
\mu {\partial I_\nu (\mu)\over \partial \tau_\nu} =
I_\nu (\mu) - \epsilon_\nu S_\nu^{\rm th} - \lambda_\nu S_\nu^{\rm Compt}(\mu) \, ,
\end{equation}
where the (generally angle-dependent) ``Compton source function'' is
given by
\begin{eqnarray}
\label{full_ang}
S_\nu^{\rm Compt}(\mu) = (1-x-6\Theta ) {\cal J}_\alpha(\mu) + 6\Theta  J +
(x+4\Theta ) {\cal J}_\beta(\mu) - 16\Theta  {\cal J}_\gamma(\mu) +
\nonumber \\
(x-3\Theta )[{\cal J}_\alpha^\prime(\mu) - {\cal J}_\beta^\prime(\mu) ] +
\Theta [{\cal J}_\alpha^{\prime \prime}(\mu)- {\cal J}_\beta^{\prime\prime}(\mu) ]
\nonumber \\
+ I_\nu(\mu) \,  b_\nu \, 2x \,
[{\cal J}_\alpha^\prime(\mu)- {\cal J}_\beta^\prime(\mu)
- {\cal J}_\alpha(\mu)+ {\cal J}_\beta(\mu)]\, ,
\end{eqnarray}
where the last term represents a contribution of the stimulated Compton
scattering, with
\begin{equation}
b_\nu \equiv {c^2 \over 2h\nu^3}\, .
\end{equation}

So far we have not made any approximations beyond the
Kompaneets limit of low photon energies ($x\ll 1$) and low electron
temperatures ($\Theta\ll 1$).  
In particular, equation (\ref{full_ang}) retains the full
angle dependence of the Compton scattering source function.  However, it
is quite complicated to solve numerically, and we have made further
approximations for all the calculations presented in this paper.

The first approximation deals with the stimulated emission term, which is
nonlinear in the specific intensity.  Fortunately, it is usually very small
compared to the spontaneous term in our models, and could be neglected
completely.  We choose to retain it, and show how to treat the nonlinearity
below.  However, we approximate the angular dependence of the radiation
field by replacing $I_\nu(\mu)$ with $J_\nu$ in the stimulated term.

More generally, we average the angle dependence of the Compton scattering
source function by first taking moments of the transfer equation 
(\ref{rtec_ang}).  Using equations (\ref{mom_alp}) - (\ref{mom_bg}), 
the first and second moments are
\begin{equation}
\label{rte_momj}
{\partial H_\nu \over \partial \tau_\nu} = 
J_\nu - \epsilon_\nu S_\nu^{\rm th} -
\lambda_\nu [ (1-x) J_\nu + (x-3\Theta )  J_\nu^\prime + 
\Theta  J_\nu^{\prime\prime} ] 
- \lambda_\nu J_\nu b_\nu 2x (J_\nu^\prime - J_\nu)\, ,
\end{equation}
\begin{equation}
\label{rte_momh}
{\partial K_\nu \over \partial \tau_\nu} = H_\nu - 
\lambda_\nu \, {2\over 5}\, \left[ (x-\Theta ) H_\nu -(x-3\Theta ) H_\nu^\prime - 
\Theta  H_\nu^{\prime\prime} \right]
-\lambda_\nu\, {2\over5}\, J_\nu b_\nu 2x(H_\nu-H_\nu^\prime) \, .
\end{equation}
We now assume that all terms on the right-hand side of equation 
(\ref{rte_momh}) are
negligible with respect to the leading term $H_\nu$, so that the second
moment equation is
\begin{equation}
\label{rte_momh2}
{\partial K_\nu \over \partial \tau_\nu} = H_\nu \, .
\end{equation}
We now introduce the variable Eddington factor,
\begin{equation}
f_\nu \equiv K_\nu / J_\nu\, ,
\end{equation}
so that the two moment equations are combined to a single equation,
\begin{equation}
\label{rte_vef}
{\partial^2 (f_\nu J_\nu) \over \partial \tau_\nu^2} = 
J_\nu - \epsilon_\nu S_\nu^{\rm th} - \lambda_\nu [ (1-x) J_\nu + 
(x-3\Theta )  J_\nu^\prime + \Theta  J_\nu^{\prime\prime} ] 
- \lambda_\nu J_\nu b_\nu 2x (J_\nu^\prime - J_\nu)\, .
\end{equation}
This equation is analogous to equations (\ref{rte}) and (\ref{compt_sf}), and
is used for computing the disk structure.

We introduce a set of discretized frequency points, $\nu_i,\, i=1,\ldots,N$;
with $\nu_i < \nu_{i+1}$.  We denote $I_i = I(\nu_i)$.
The frequency derivatives are treated using the following discrete
representation. Let us first take internal frequency points,
$i=2,\ldots, N-1$. Then
\begin{equation}
\label{iprim1}
I_i^\prime = c_i^{-} I_{i-1} + c_i^{0} I_{i} + c_i^{+} I_{i+1} \, ,
\end{equation} 
\begin{equation}
\label{iprim2}
I_i^{\prime\prime} = d_i^{-} I_{i-1} + d_i^{0} I_{i} + d_i^{+} I_{i+1} \, ,
\end{equation} 
where the coefficients $c_i^{-}$,  $c_i^{0}$,  $c_i^{+}$ are
determined by the method of Chang \& Cooper (1970), and
\begin{equation}
d_i^{-} =  {2 \over \Delta_{i-1/2}\Delta_{i} }\, ,
\end{equation} 
\begin{equation}
d_i^{+} =  {2 \over \Delta_{i+1/2}\Delta_{i} }\, ,
\end{equation} 
\begin{equation}
d_i^{0} = - d_i^{-} - d_i^{+}\, ,
\end{equation} 
where
\begin{equation}
\Delta_{i-1/2} = \ln(\nu_{i}/\nu_{i-1})\, , \quad\quad
\Delta_{i+1/2} = \ln(\nu_{i+1}/\nu_i)\, ,
\end{equation} 
and
\begin{equation}
\Delta_{i} = \Delta_{i-1/2} + \Delta_{i+1/2}\, .
\end{equation} 
The boundary conditions in frequency space are written in the form
of equations (\ref{iprim1}) and ({\ref{iprim2}), with
$c_1^- = d_1^- = c_1^+ = d_1^+ = 0$, and 
$c_N^- = d_N^- = c_N^+ = d_N^+ = 0$.  
At the lowest frequency we have, assuming the Rayleigh-Jeans form of the
specific intensity, equation (\ref{jlbound}),
\begin{equation}
c_1^0 = 2, \quad\quad d_1^0 = 4\, .
\end{equation} 
At the highest frequency we have either
\begin{equation}
c_N^0 = 3-z, \quad\quad d_N^0 = (3-z)^2 - z \, ,
\end{equation} 
(with $z=h\nu/kT$) if we assume the Wien form of the specific intensity,
equation (\ref{jubound}); or
\begin{equation}
c_N^0 = -\beta, \quad\quad d_N^0 = \beta^2 \, ,
\end{equation} 
if we assume the power-law form, equation (\ref{jupl}).

We write the radiative transfer equation for the $i$-th frequency point as
\begin{equation}
\label{rte_i}
{\partial^2 (f_i J_i) \over \partial \tau_i^2} = 
J_i - \epsilon_i S_i^{\rm th} - \lambda_i [
\left( A_i J_{i-1} + B_i J_i + C_i J_{i+1} \right)
- J_i \left(E_i J_i + U_i J_{i-1} + V_i J_{i+1} \right) ]\, ,
\end{equation}
where
\begin{equation}
B_i = (1-x_i) + (x_i - 3\Theta ) c_i^0 + \Theta  d_i^0 \, ,
\end{equation}
\begin{equation}
A_i = (x_i - 3\Theta ) c_i^- + \Theta  d_i^- \, ,
\end{equation}
\begin{equation}
C_i = (x_i - 3\Theta ) c_i^+ + \Theta  d_i^+ \, ,
\end{equation}
\begin{equation}
E_i = b_i 2x_i (c_i^0 -1)\, ,
\end{equation}
\begin{equation}
U_i = b_i 2x_i c_i^- \, ,
\end{equation}
\begin{equation}
V_i = b_i 2x_i c_i^+ \, .
\end{equation}
We now avoid the nonlinearity of the stimulated emission term by evaluating
it partly by using ``old'' mean intensities.
In other words, we replace $B_i$ by $B_i^\prime$, where
\begin{equation}
B_i^\prime = B_i + E_i J_i^{\rm old} + U_i J_{i-1}^{\rm old} + 
V_i J_{i+1}^{\rm old} \, .
\end{equation}
The discretized transfer equation is then written
\begin{equation}
\label{rteapp_ii}
{\partial^2 (f_i J_i) \over \partial \tau_i^2} = 
J_i - \epsilon_i S_i^{\rm th} - \lambda_i
\left( A_i J_{i-1} + B_i^\prime J_i + C_i J_{i+1} \right)\, ,
\end{equation}
The final step is to consider a discrete representation of the differential
term as in the standard Feautrier method (Mihalas 1978), viz.
\begin{equation}
{\partial^2 (f_i J_i) \over \partial \tau_i^2} \rightarrow
\alpha_{i,d} J_{i,d-1} - \beta_{i,d} J_{i,d} + \gamma_{i,d} J_{i,d+1}\, ,
\end{equation}
where index $d$ labels depth points, while $i$ labels the frequency
points. The coefficients $\alpha$, $\beta$, and $\gamma$ (not to be
confused with the previously introduced angle-dependent matrices of
Compton scattering) are given by
\begin{equation}
\alpha_{i,d} = f_{i,d-1}/(\Delta\tau_{i,d-1/2} \Delta\tau_{i,d} )\, ,
\end{equation}
\begin{equation}
\gamma_{i,d} = f_{i,d+1}/(\Delta\tau_{i,d+1/2} \Delta\tau_{i,d} )\, ,
\end{equation}
\begin{equation}
\beta_{i,d} = {f_{i,d}\over \Delta\tau_{i,d} }
\left({1 \over \Delta\tau_{i,d-1/2} } + {1 \over \Delta\tau_{i,d+1/2}}
\right)\, ,
\end{equation}
and
\begin{equation}
\Delta\tau_{i,d-1/2} = \tau_{i,d} - \tau_{i,d-1}\, ,
\end{equation}
\begin{equation}
\Delta\tau_{i,d+1/2} = \tau_{i,d+1} - \tau_{i,d}\, ,
\end{equation}
\begin{equation}
\Delta\tau_{i,d} = (\tau_{i,d+1/2} + \tau_{i,d-1/2})/2 \, .
\end{equation}
The (second-order) boundary conditions are given by
\begin{equation}
\beta_{i,1} = 2 f_{i,1}^H/\Delta\tau_{i,3/2} + 2 
f_{i,1}/(\Delta\tau_{i,3/2})^2\, ,
\end{equation}
\begin{equation}
\gamma_{i,1} = 2 f_{i,2}/(\Delta\tau_{i,3/2})^2\, ,
\end{equation}
and 
\begin{equation}
\beta_{i,D} =  2f_{i,D}/(\Delta\tau_{i,D-1/2})^2\, ,
\end{equation}
\begin{equation}
\alpha_{i,D} = 2f_{i,D-1}/(\Delta\tau_{i,D-1/2})^2\, .
\end{equation}

The final discretized form of the transfer equation thus reads
\begin{equation}
\label{rte_discr}
-\alpha_{i,d} J_{i,d-1} - \gamma_{i,d} J_{i,d+1} +
(\beta_{i,d} + 1 - \lambda_{i,d} B_{i,d}^\prime ) J_{i,d} 
-\lambda_{i,d}  A_{i,d} J_{i-1,d} - \lambda_{i,d} C_{i,d} J_{i+1,d} = 
\epsilon_{i,d} S_{i,d}^{\rm th} \, .
\end{equation}
The formal solution, i.e., a solution of the transfer equation 
with the known thermal source function, $S_{id}^{\rm th}$,
is obtained as follows.
We organize the mean intensities in a set of column vectors
\begin{equation}
{\bf J}_i \equiv \left( J_{i,1}, J_{i,2}, \ldots, J_{i,D} \right)^T \, ,
\end{equation}
so that the mean intensity vector contains intensities in all depth points
for a given frequency point $i$. The resulting discretized transfer equation
reads
\begin{equation}
\label{rte_mat1a}
- {\bf A}_i {\bf J}_{i-1} + {\bf B}_i {\bf J}_i - {\bf C}_i {\bf J}_{i+1} =
{\bf L}_i\, ,
\end{equation}
where the matrices ${\bf B}$ are tridiagonal (because of the difference 
representations of the second derivative with respect to depth), 
while the matrices ${\bf A}$ and ${\bf C}$ are diagonal (because the terms
containing the frequency derivatives are local in physical space).

The matrix elements are given by
\begin{equation}
({\bf B}_i)_{d,d} = \beta_{i,d} +1 - \lambda_{i,d} B_{i,d}^\prime \, , 
\end{equation}
\begin{equation}
({\bf B}_i)_{d,d-1} = - \alpha_{i,d}\, , 
\end{equation}
\begin{equation}
({\bf B}_i)_{d,d+1} = - \gamma_{i,d}\, , 
\end{equation}
\begin{equation}
({\bf A}_i)_{d,d} = \lambda_{i,d} A_{i,d}\, , 
\end{equation}
\begin{equation}
({\bf C}_i)_{d,d} = \lambda_{i,d} C_{i,d}\, , 
\end{equation}
\begin{equation}
({\bf L}_i)_{d} = \epsilon_{i,d} S_{i,d}^{\rm th}\, . 
\end{equation}
Solution of equation (\ref{rte_mat1a}) is done by the
standard Gauss-Jordan elimination, consisting of a forward elimination 
followed by a backsubstitution, 
\begin{equation}
\label{felimb}
{\bf D}_i = ({\bf B}_i - {\bf A}_i {\bf D}_{i-1})^{-1} \, 
{\bf C}_i \, ,\quad\quad i=2,\ldots, D\, ,
\end{equation}
with
\begin{equation}
{\bf D}_1 = {\bf B}_1^{-1} \, {\bf C}_1 \, ,
\end{equation}
and
\begin{equation}
\label{e6--41}
{\bf Z}_i = ({\bf B}_i - {\bf A}_i {\bf D}_{i-1})^{-1} \, 
({\bf L}_i + {\bf A}_i {\bf Z}_{i-1})\, , \quad\quad
i=2,\ldots, D\, ,
\end{equation}
with
\begin{equation}
{\bf Z}_1 = {\bf B}_1^{-1} \, {\bf L}_1 \, .
\end{equation}
The back substitution step is
\begin{equation}
\label{felime}
{\bf J}_i = {\bf D}_i {\bf J}_{i+1} + {\bf Z}_i\, \quad\quad i=1,\ldots, D-1\, ,
\end{equation}
with ${\bf J}_{N}=Z_N$.
Starting at $i=1$, we compute successive values for ${\bf D}_d$ 
and ${\bf Z}_i$ through
$i=N-1$. At the last point, $i=N$, ${\bf C}_N\equiv 0$ and hence ${\bf D}_N=0$, 
and ${\bf J}_N = {\bf Z}_N$.
Having found ${\bf J}_N$, we then perform successive back-substitutions into
equation (\ref{felime}) to find ${\bf J}_i, \, i=N-1,\ldots, 1$.

\section{Appendix B: Treatment of Dielectronic Recombination}

In terms of a cross-section, the photoionization rate is given by
(see, e.g., Mihalas 1978)
\begin{equation}
\label{rik}
R_{ik} = \int_{\nu_0}^\infty {4\pi\over h\nu} \sigma(\nu)  J_\nu d\nu\, ,
\end{equation}
where $\sigma(\nu)$ is the photoionization cross-section, $\nu$ the
frequency, $\nu_0$ the frequency of the continuum edge, and $J_\nu$ the 
mean intensity of radiation.  The subscript
$i$ labels the starting level of the bound-free transition, while $k$
labels the ending level (usually the ground state of the next higher ion).

The radiative recombination rate is given through detailed balance
arguments (the Einstein-Milne relations for the continuum; see also
Mihalas 1978) as
\begin{equation}
\label{rki}
R_{ki} = n_{\rm e} \phi_i(T) \int_{\nu_0}^\infty {4\pi\over h\nu} \sigma(\nu)  
\left({2h\nu^3\over c^2}  + J_\nu \right) \exp(-h\nu/kT) d\nu\, ,
\end{equation}
where $n_{\rm e}$ is the electron density, and $\phi_i(T)$ the 
Saha-Boltzmann factor, given by  (Mihalas 1978, p. 113)
\begin{equation}
\label{sbf}
\phi_i(T) = \left( {h^2 \over 2\pi m k T}\right)^{3/2}\, 
{g_i \over 2 g_k} \exp(h\nu_0/ k T) 
\equiv C_I T^{-3/2} \, {g_i \over g_k}\, \exp(h\nu_0/ k T) \, ,
\end{equation}
where $g$ are the statistical weights.
The first term on the right-hand side of equation (\ref{rki}) represents
spontaneous recombination, while the second term (proportional to 
$J_\nu$) represents stimulated recombination.

We now use this to express the known dielectronic recombination rate through
an artificial cross-section.  For simplicity, we assume that
\begin{equation}
\sigma(\nu) = \sigma_0\, , \quad{\rm for}\quad \nu_0 \leq \nu \leq \nu_1\, ,
\end{equation}
and is zero everywhere else. We further assume that $\nu_1/\nu_0 - 1 \ll 1$,
i.e., the frequency $\nu_1$ is not very different from $\nu_0$.
We further assume that the stimulated term is negligible
(i.e., $J_\nu \ll 2h\nu^3/c^2$).  The known recombination rate is given
through the artificial cross-section as follows
\begin{equation}
R = n_{\rm e}\, C_I\, T^{-3/2} \, {g_i \over g_k}\,\exp(h\nu_0/ k T)\,
\int_{\nu_0}^{\nu_1} {4\pi\over h\nu} \sigma_0  
{2h\nu^3\over c^2} \exp( -h\nu/ k T) \, ,
\end{equation}
or 
\begin{equation}
R = C_I {8\pi \over c^2} n_{\rm e}\, T^{-3/2} \, {g_i \over g_k}\, \sigma_0
\int_{\nu_0}^{\nu_1} \nu^2\, \exp [-h(\nu-\nu_0)/ k T] \, .
\end{equation}
The integrand is the product of two terms. The first one, $\nu^2$, varies
slowly with $\nu$ compared to the second, exponential, term.
We may therefore remove the $\nu^2$ term from the integral, replacing it
with $\bar \nu^2 \equiv [(\nu_0 + \nu_1)/2]^2 \simeq\nu_0^2$, 
and performing the remaining integration analytically,
\begin{equation}
R = C_I \, {8\pi k\over h c^2}\, n_{\rm e} \,T^{-1/2} 
\, {g_i \over g_k}\,\sigma_0 \, \bar\nu^2
\left[ 1 - \exp(-h \Delta\nu/ kT) \right]\, ,
\end{equation}
where $\Delta\nu = \nu_1 - \nu_0$.
In the calculations reported in this paper, we used  
$\nu_1 = 1.1 \nu_0$.

Consequently, the desired value of the cross-section, $\sigma_0$, is given by
\begin{equation}
\sigma_0 = R {\sqrt{T}\over n_{\rm e}} \, C_I\, {h c^2 \over 8 \pi k}\,
{g_k \over g_i}\, 
{1\over \bar\nu^2}\, \left[ 1 - \exp\left(-{h \Delta\nu/ kT}\right) 
\right]^{-1}\, .
\end{equation}
Unlike the traditional cross-section, $\sigma_0$ now generally 
depends on temperature and electron density.  We treat this 
dependence exactly.
Nevertheless, the dependence is rather weak because the
dielectronic recombination rate is typically proportional to $T^{-1/2}$, so
the $T$-dependence is contained essentially in the term
$\exp[-h(\nu_1-\nu_0)/kT]$, which is typically small compared to unity.
Analogously, the dielectronic recombination rate is directly proportional
to $n_{\rm e}$, so the dependence of $\sigma_0$ on physical parameters
is to a large extent factored out, and one is left with $\sigma_0$
essentially independent of state parameters.

\end{document}